%% file: main.tex
\documentclass[conference]{IEEEtran}
\IEEEoverridecommandlockouts
\usepackage{cite}
\usepackage{amsmath,amssymb,amsfonts}
\usepackage{algorithmic}
\usepackage{graphicx}
\usepackage{textcomp}
\usepackage{xcolor}
\usepackage{subcaption}  
\def\BibTeX{{\rm B\kern-.05em{\sc i\kern-.025em b}\kern-.08em
    T\kern-.1667em\lower.7ex\hbox{E}\kern-.125emX}}
\begin{document}

\title{\textit{Glitch in the Sky}: Exploiting Voltage Fault Injection in UAV Flight Controllers\\
}

\author{\IEEEauthorblockN{Yun-Ping Hsiao\IEEEauthorrefmark{1}\footnotemark{$\dagger$},
Yanda Li\IEEEauthorrefmark{1}, Youssef Gamal\IEEEauthorrefmark{1}, Halima Bouzidi\IEEEauthorrefmark{1},  Mohammad Abudllah Al Faruque\IEEEauthorrefmark{1}}
\IEEEauthorrefmark{1}Dept. of Electrical Engineering and Computer Science, University of California, Irvine, CA, USA \\
\IEEEauthorrefmark{1}\{yunpinh, yandal5, yhmahmou, hbouzidi, alfaruqu\}@uci.edu
\\
}

\maketitle

\footnotetext[1]{$\dagger$ Corresponding Author.}

\input{sections/abstract}

\input{sections/introduction}

\input{sections/background}

\input{sections/threat_model}

\input{sections/approach}

\input{sections/evaluation}

\input{sections/conclusion}

\bibliographystyle{IEEEtran}
\bibliography{bibliography}

\end{document}

%% file: sections/abstract.tex
\begin{abstract}
As Cyber-Physical Systems (CPS) become increasingly pervasive and autonomous, ensuring the resilience of their embedded logic is critical to maintaining safety and integrity. Among the most stealthy and damaging threats are non-invasive fault injection attacks, where hardware-level disturbances propagate into software execution and compromise control logic. In this paper, we investigate the susceptibility of Unmanned Aerial Vehicle (UAV) autopilot fail-safe mechanisms to voltage glitch fault injection. We introduce a dual evaluation approach: software-based fault simulation using ARMORY and hardware-based experiments with a voltage glitching platform (Chip-Whisperer), applying controlled and timely faults to an STM32 microcontroller running UAV-Autopilot fail-safe logic. Our targeted analysis of specific fail-safe modes uncovers timing-sensitive vulnerabilities that can suppress or alter safety responses, such as disabling emergency failsafe activation at critical moments, potentially enabling UAV hijacking. Furthermore, we validate software-based fault injection results against real hardware behavior, demonstrating how simulated attacks translate into tangible risks for CPS security and reliability.
\end{abstract}

\begin{IEEEkeywords}
Hardware faults, Voltage glitching, Cyber-Physical Systems
\end{IEEEkeywords}

%% file: sections/introduction.tex
\section{Introduction}
Fault injection \cite{hsueh1997fault} is a powerful evaluation technique used to induce unintended behavior in a Device Under Test (DUT) by temporarily disturbing its runtime environment, either at the software level \cite{mortlock2024adaptivedatafusion, Sargolzaei2018} or through physical hardware manipulation. Hardware-level fault injection methods range from clock glitching and voltage fault injection (VFI) to electromagnetic (EM) radiation and laser fault injection. These techniques momentarily disrupt the normal operation of a device. The potential consequences are named as Fault Model, including bit flip \cite{rakin2021t}, skipping of machine instructions \cite{timmers2017escalating} or altering the data stored on a device’s internal memory \cite{flipbit2014}. 

To date, much of the fault injection research has focused on security-critical operations, such as cryptographic implementations (e.g., AES, RSA) ~\cite{barenghi2009low, mishra2025modern, yang2020design, moro2013electromagnetic} and secure boot mechanisms \cite{Fanjas2023}. These studies have demonstrated that carefully timed faults can bypass firmware verification, extract secret keys, or undermine trusted execution environments. On the other sides, the broader field of Cyber-Physical Systems (CPS) security \cite{wan2015securityaware, faezi2019oligo} has gained significant traction, as researchers demonstrate how cyber-attacks can manipulate physical outcomes \cite{wan2018physical, chhetri2018information, chhetri2017fix}.  However, the security implications of fault injection on Cyber-Physical Systems (CPS) \cite{chhetri2017security, lopez2017security}, such as Unmanned Aerial Vehicles (UAVs), has been understudied, especially at high-level application logic \cite{chhetri2017cross, chhetri2016thermal, rashid2017modeling, balaji2015models}. In these systems, safety and reliability are paramount, and faults in control logic can lead to catastrophic outcomes (e.g., UAV hijacking, mission failure, or even crashes) \cite{al2022cross, xie2025flytrap}.

Fault injection testing plays a vital role in evaluating the resilience of UAV systems. One of the most widely used open-source autopilot platforms, PX4-Autopilot, includes safety-critical components such as failsafe mechanisms designed to prevent accidents when sensor errors or communication failures occur. These routines are foundational to drone safety in autonomous or semi-autonomous flight.

However, fault injection poses a serious security and safety risk to UAVs. An attacker is capable of injecting faults when the failsafe logic is executing, resulting in uncontrolled behavior or crash scenarios. As modern drones become more connected and complex, their exposure to both remote and physical attacks grows. Despite this risk, no prior work has comprehensively investigated hardware fault injection on PX4 or other CPS platforms, leaving a critical gap in CPS system security research \cite{chang2018survey, RAMPO}.

This paper aims to bridge that gap by investigating the vulnerability of one of the widely used autopilots for UAVs, PX4-Autopilot, and its safety mechanisms to fault injection attacks. To achieve this, the study will employ a dual approach: conducting software-based simulation using ARMORY and performing hardware-based fault injection with ChipWhisperer paired with an STM32 microcontroller. The main objectives are to evaluate the effectiveness and characterize the consequences of these fault injections on system behavior.

This work makes several contributions: 
\begin{itemize}
    \item It addresses a critical gap in Cyber-Physical Systems (CPS) security research by examining hardware-level vulnerabilities, which are often overlooked compared to well-studied software-level threats.
    \item A dual-method fault injection campaign targeting PX4-Autopilot, combining simulation-based attacks using ARMORY with physical hardware glitching on STM32 microcontrollers.
    \item A security evaluation of PX4’s failsafe routines, identifying timing-sensitive attack windows and characterizing the impact of injected faults on decision logic.
    \item Demonstration of the complementary strengths of simulation and hardware testing, bridging the gap between abstract fault models and real-world hardware behavior.
\end{itemize}

%% file: sections/background.tex
\section{Background and Related Work}

\subsection{Unmanned Aerial Vehicles Autopilot Security}
Unmanned Aerial Vehicles (UAVs) are increasingly deployed in fields ranging from agriculture and logistics to surveillance and defense. Many of these systems rely on open-source autopilot frameworks, with PX4 being one of the most widely adopted due to its modular architecture and support for diverse platforms. The growing autonomy and connectivity of UAVs are accompanied by escalating cybersecurity vulnerabilities. Existing UAV security research has primarily focused on software-level threats. Communication protocols like MAVLink, used extensively by PX4 for telemetry and control, have been shown to be susceptible to injection and denial-of-service (DoS) attacks~\cite{hsu2021ids}. Furthermore, software fuzzing has uncovered critical vulnerabilities where malformed messages can crash or hijack UAV behavior. ~\cite{yu2023cybersecurity, kim2021pgfuzz, amorim2025enforcingmavlinksafety, malviya2024fuzzing}. Navigation security is another well-studied area, with research focusing on defending against GPS spoofing by enhancing the Extended Kalman Filter (EKF) with anomaly detection. ~\cite{jung2024gps}
While this software-centric research is valuable, it often overlooks threats at the hardware level. PX4-based UAVs typically run on resource-constrained microcontrollers that lack strong physical protections. These platforms remain exposed to Fault Injection (FI) attacks~\cite{hsueh1997fault} , which can silently disrupt control logic or bypass safety mechanisms without triggering software-level alarms. Despite their potential severity, hardware-based fault injection remains significantly underexplored in the context of UAV autopilot resilience.

\begin{figure}[h!]
    \centering
    \includegraphics[width=.47\textwidth]{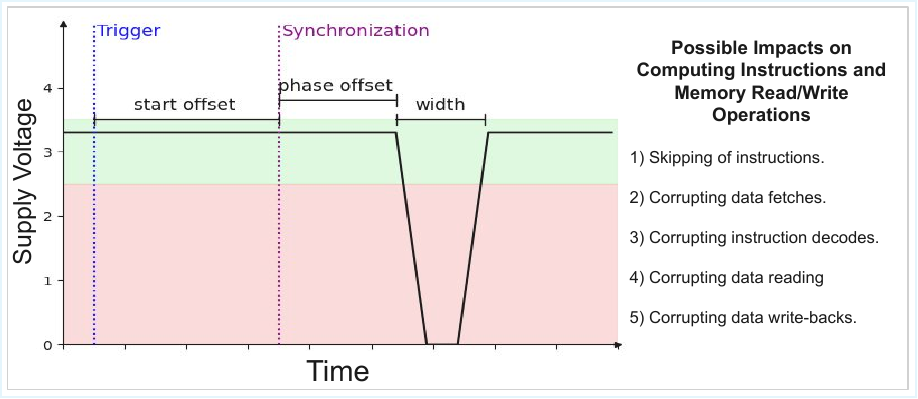} 
    \caption{A voltage fault injection attack and its potential impacts \cite{Multi-Glitch_2023}}
    \label{fig:voltage_glitch}
\end{figure}

\subsection{Hardware Fault Injection}
Fault injection is a powerful technique used to evaluate the robustness of embedded systems or to compromise their security. It involves the deliberate introduction of transient faults into a device's physical environment during runtime to induce unintended behaviors. These attacks have been widely applied to compromise cryptographic hardware, such as extracting secret keys from AES or RSA implementations ~\cite{barenghi2009low, mishra2025modern}, and to undermine secure boot mechanisms in microcontrollers and IoT devices~\cite{korak2014effects, OFlynn2016FaultIU}.
Faults can be induced through various environmental manipulations, including clock glitching, electromagnetic interference, and laser-based injection ~\cite{breier2022practical}. These disturbances typically result in a Fault Model, which describes the logical outcome of the physical attack. Common fault models include bit-flips in memory, the corruption of data stored in registers, or the skipping of critical machine instructions. When targeted at control-flow logic, these faults allow an attacker to redirect program execution, effectively bypassing security checks or safety routines.

\subsection{Voltage-based Fault Injection}
Among the various methods of fault injection, Voltage Glitching, often referred to as a Power Fault, is frequently favored due to its lower cost and effectiveness across a wide range of embedded targets ~\cite{breier2022practical}. It involves introducing a short, intentional disturbance or "glitch" into the power supply ($\text{V}_{CC}$) of a processor. This brief voltage drop disrupts the normal timing of digital logic gates, leading to effects such as instruction skips or incorrect branching.
The success of a power fault depends on the critical relationship between voltage stability and the precise timing of digital operations. As illustrated in Figure~\ref{fig:voltage_glitch}, an effective glitch requires both a sufficient voltage drop and carefully controlled timing. The following parameters ~\cite{OFlynn2016FaultIU, Multi-Glitch_2023} are essential for configuring a successful voltage glitch:

\begin{itemize} 
\item \textbf{Trigger:} The external event or signal that initiates the glitching sequence. This serves as the temporal starting point for the attack.
\item \textbf{Start Offset:} The time delay, usually measured in clock cycles, between the \textit{Trigger} and  the \textit{Synchronization} point. 
\item \textbf{Synchronization:} A specific reference point in time within the target system's execution used to align the fault with a specific instruction.
\item \textbf{Phase Offset:} A fine-grained delay used to shift the glitch relative to the clock edge. A positive offset occurs after the rising edge, while a negative offset occurs before it.
\item \textbf{Width:} The duration of the voltage drop. This determines the overall impact on the target's logic gates and whether the fault results in a reset or a successful glitch.
\end{itemize}

By precisely timing these parameters to coincide with critical decision-making logic, such as the PX4 failsafe routines, an attacker can fundamentally subvert the drone's intended safety state.
In Figure \ref{fig:voltage_glitch}, the green area represents the normal operating voltage range of the device, while the red shaded area indicates the voltage levels where the operation of device may become unreliable or faulty. The voltage level that separates these ranges is device-specific and depends on factors such as the architecture of microcontroller, manufacturing process, and operating frequency.

\subsection{PX4-Autopilot} 
\begin{figure}[ht]
    \centering
    \includegraphics[width=0.45\textwidth]{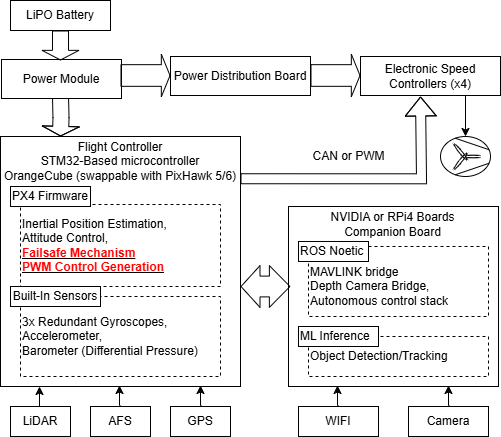}
    \caption{Architecture of a PX4-based UAV system showing the interaction between the STM32-based flight controller and the companion board. The Failsafe Mechanism  occur within the PX4 firmware layer on the STM32 microcontroller.}
    \label{fig:px4_uav_architecture}
\end{figure}

PX4 is an open-source, professional-grade flight control system designed for a wide range of unmanned aerial vehicles (UAVs) \cite{px4_ieee}. As illustrated in Figure~\ref{fig:px4_uav_architecture}, the PX4 firmware runs on an STM32-based flight controller. This controller interfaces with onboard sensors, communicates with a companion computer (such as an NVIDIA or Raspberry Pi board), and controls the drone’s motors through PWM or CAN protocols. The failsafe mechanism is embedded directly within the flight control logic as a critical part of the firmware.

\subsection{Failsafe Mechanism: An Overview}
The Failsafe logic within PX4 serves as the final line of defense for flight safety during degraded system conditions \cite{px4_safety}. It monitors key inputs; such as radio connectivity, GPS validity, and battery levels, and executes recovery actions to prevent catastrophic crashes. By default, PX4 enters a temporary Hold mode to allow for pilot intervention; if no override occurs, the system executes a final safety action. These actions are categorized by severity and include \textit{Return-to-Launch (RTL)}, \textit{Land}, \textit{Disarm}, or \textit{Flight Termination}, among others. Table~\ref{tab:failsafe-actions} introduces the possible failsafe actions defined in PX4 documentation, ordered by increasing severity.

\begin{table*}
\centering
\caption{Failsafe Action Modes}
\label{tab:failsafe-actions}
\scalebox{1.0}{
\begin{tabular}{l|l|l} 
\hline
\multicolumn{1}{c|}{\textbf{Action Mode}} & \multicolumn{1}{c|}{\textbf{Action Mode Code}} & \multicolumn{1}{c}{\textbf{Description}} \\ 
\hline
None/Disable & action = 0 & No response is triggered; the failsafe is ignored. \\
Warning & action = 1 & Only a warning message is issued (e.g., via QGroundControl). \\
Hold & action = 5 & The vehicle holds position (Multicopter hovers; Fixed-Wing~circles). \\
Return-to-Launch (RTL) & action = 6 & The vehicle returns to the launch point using a safe, preplanned~path. \\
Land & action = 7 & The vehicle lands at the current or designated safe location. \\
Disarm & action = 9 & The motors are stopped immediately. \\
Flight Termination & action = 10 & All controllers are disabled; PWM outputs are set to failsafe~values (e.g., for parachute deployment). \\
\hline
\end{tabular}
}
\end{table*}

The effectiveness of these mechanisms relies entirely on the integrity of the microcontroller's execution flow. However, source-level analysis of PX4 reveals that the logic governing these states is complex and includes undocumented behaviors, such as internal \textit{Fallback} modes and specific \textit{ActionOptions} attributes. Because these routines rely on simple conditional checks at the machine-code level, they are highly susceptible to the power faults described previously. An attacker targeting the "decision window" of a failsafe check could force the system to ignore a critical sensor failure, leading to a mission failure or a crash.

%% file: sections/threat_model.tex
\section{Threat Model}

\subsection{Attack Scenario: The Black-Box Assumption}
A core element of this threat model is its operational stealth, which leverages the physical enclosure of the hardware as a security blind spot. Commercial flight controllers are typically treated as black-box appliances by both end-users and system integrators.
The complex, multi-stage nature of an untrusted supply chain; spanning manufacturers, distributors, and retailers, provides an attacker with multiple entry points. During transit, an attacker can intercept the device, remove its external enclosure to install the hardware implant, and seamlessly reseal the shell before the product ever reaches the customer.

Because the malicious PCB is installed inside the housing prior to final delivery, draws parasitic power, and remains physically concealed within internal voids, it is visually undetectable. Standard operational procedures, visual inspections, and routine pre-flight diagnostic checks do not require opening the factory-sealed shell. Consequently, the physical enclosure acts as an effective shield, allowing the discrete hardware modification to remain persistent and entirely hidden throughout the lifecycle of the drone.

\subsection{Attacker Capabilities and Assumptions}
We assume an attacker capable of designing and fabricating miniaturized, application-specific printed circuit boards (PCBs). This separate, malicious board is functionally equivalent to specialized fault injection tooling (such as a localized ChipWhisperer), but is explicitly optimized for size, autonomy, and in-situ deployment within the target's physical enclosure rather than a controlled laboratory environment.

Furthermore, we assume the attacker procures an identical reference flight controller to conduct offline profiling. This allows them to experimentally map the target's execution flow and determine the precise trigger condition, the glitch offset, and the glitch width required for a successful fault injection prior to finalizing the implant's design.

\subsection{Attack Vector}
The primary attack mechanism is a discrete, secondary PCB physically implanted into the flight controller's housing. To bypass the spatial constraints of the highly confined internal volume, the attacker employs integration techniques such as Corner Tucking (exploiting small spatial voids adjacent to structural mounting standoffs) or Component Swapping (replacing benign passive components with the miniaturized flex-PCB payload).

To execute the attack autonomously, the discrete malicious PCB must harvest power directly from the target board’s primary voltage rails. However, because the attack relies on momentarily shorting this shared primary rail to induce the necessary voltage drop for the glitch, the implant requires a self-preservation mechanism. A dedicated filtering and capacitive isolation stage, functioning as a localized energy reservoir, is implemented on the malicious PCB. This ensures its own microcontroller remains stable and immune to the sudden voltage collapse it intentionally generates against the target.

%% file: sections/approach.tex
\section{Approach}
\begin{figure}[ht]
    \centering
    \includegraphics[width=\linewidth]{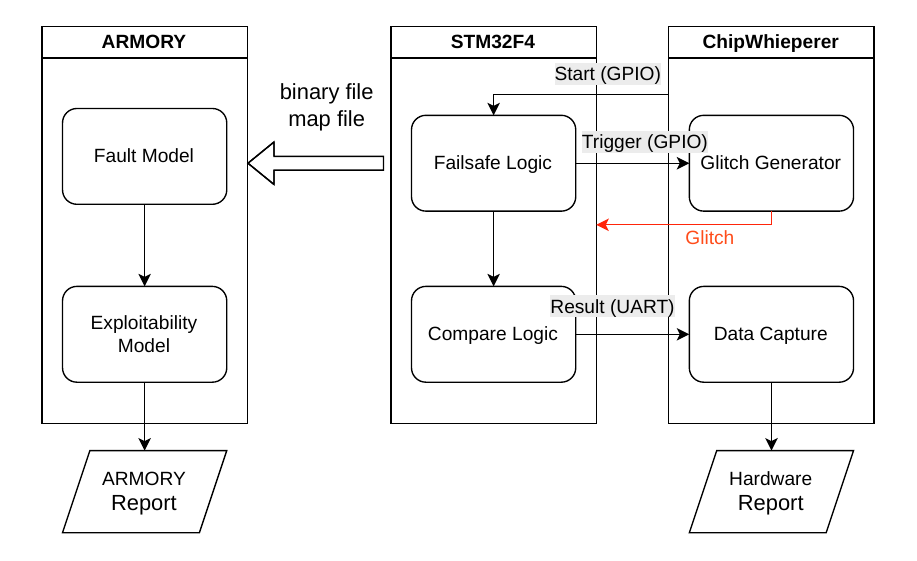}
    \caption{Integrated simulation workflow.}
    \label{fig:simulation workflow}
\end{figure}

To systematically evaluate how power faults propagate through the UAV system, we utilize a two-phase methodology: software simulation to analyze the vulnerability of functions at the machine-code level, followed by hardware-based fault injection to validate the physical exploitability. Figure \ref{fig:simulation workflow} illustrates the integrated workflow.

ARMORY~\cite{ARMORY}, a software-based fault injection framework, provides broad coverage of fault models, high repeatability of test cases, and detailed insight into the functional impact of faults without requiring physical hardware instrumentation. The analyzed source code can be evaluated directly as a compiled binary executing on the STM32 architecture. Following the simulation, hardware-based fault injection is performed on an STM32 microcontroller using precise voltage glitching. The experiment utilizes an intact STM32 target board, configured non-destructively without removing decoupling capacitors, and a ChipWhisperer board~\cite{OFlynn2016FaultIU,oflynn_blog,oflynn_video}. This dedicated, autonomous flow allows us to iterate through multiple glitch parameter sets for comprehensive downstream analysis.

\subsection{Software Simulation}
ARMORY is an open-source, automated framework designed to perform fault simulation on ARM-M binaries. It emulates the behavior of ARM processors, allowing for the injection of various fault models at the machine-code level to analyze their impact on program execution. Because the STM32 microcontroller, commonly used in PX4 flight controllers, is based on the ARM Cortex-M4 architecture, ARMORY serves as an ideal simulation environment for this analysis.

All transient fault models provided by the ARMORY framework were utilized to comprehensively map vulnerabilities. These models are categorized into instruction faults and register faults, each designed to represent specific hardware-induced errors:

\begin{itemize} 
\item \textbf{Instruction Faults: }
This category includes Instruction Skip, Instruction Byte-Set, Instruction Byte-Clear, and Instruction Bit-Flip. These models simulate interruptions that alter the binary encoding of an instruction during execution, resulting in forced changes to control flow, computation logic, or data handling.
\item \textbf{Register Faults: }
 These models simulate corruption within CPU register values, which can critically affect program logic and memory operations. Two specific temporal categories were explored: transient faults and until-overwrite faults. Transient register faults represent temporary corruption affecting only the immediate instruction utilizing the faulty value, after which the register restores to its original state. Until-overwrite faults persist in the register until explicitly updated by a subsequent instruction. Specific fault types injected include Register Clear, Register Fill, Register Byte-Set, Register Byte-Clear, and Register Bit-Flip.
\end{itemize}

In ARMORY simulations, synchronization is achieved by defining symbolic execution windows based on specific code regions of interest. Rather than relying on real-time hardware signals, ARMORY uses manually assigned start and halt symbols extracted from the disassembly output of the target binary. For all simulations in this study, the start point was consistently defined as the start symbol of the targeted helper function relevant to the scenario under test.

\subsection{Hardware Characterization}
The ChipWhisperer is an open-source device designed for conducting fault injection attacks on embedded systems. It is equipped with a dedicated glitch module and a MOSFET-based crowbar circuit that connects directly to the target device's $V_{DD}$ line to induce a voltage drop. Because the experiment was designed to be non-destructive (i.e., retaining the target board's factory decoupling capacitors), a "longer" glitch width was required to successfully pull the voltage down compared to default ChipWhisperer setups. To achieve this, we configured the ChipWhisperer to \texttt{"enable-only"} mode, which activates the glitch circuit for entire clock cycles rather than sub-cycle fractions. The ChipWhisperer's frequency was configured to 32 MHz, utilizing the \textit{External Offset} and \textit{Repeat Count} parameters to precisely control the fault timing and duration.

Flight controller boards directly manage essential physical operations, making them high-impact targets for security analysis. Within the PX4 ecosystem, STM32 microcontrollers (specifically the STM32H7 and STM32F series) are the most prevalent architectures used for main processors and failsafe co-processors. To accurately model this environment while maintaining experimental control, the STM32F407-Discovery board was selected as the primary hardware platform. It offers a lower level of abstraction than a fully assembled flight controller, providing direct access to the microcontroller’s pins for reliable synchronization. To further isolate the target function and ensure the base program ran predictably during profiling, the PX4 API layer was temporarily removed, simplifying the execution environment.

\subsection{Experiment Setup}
\begin{figure}[ht]
    \includegraphics[width=\linewidth]{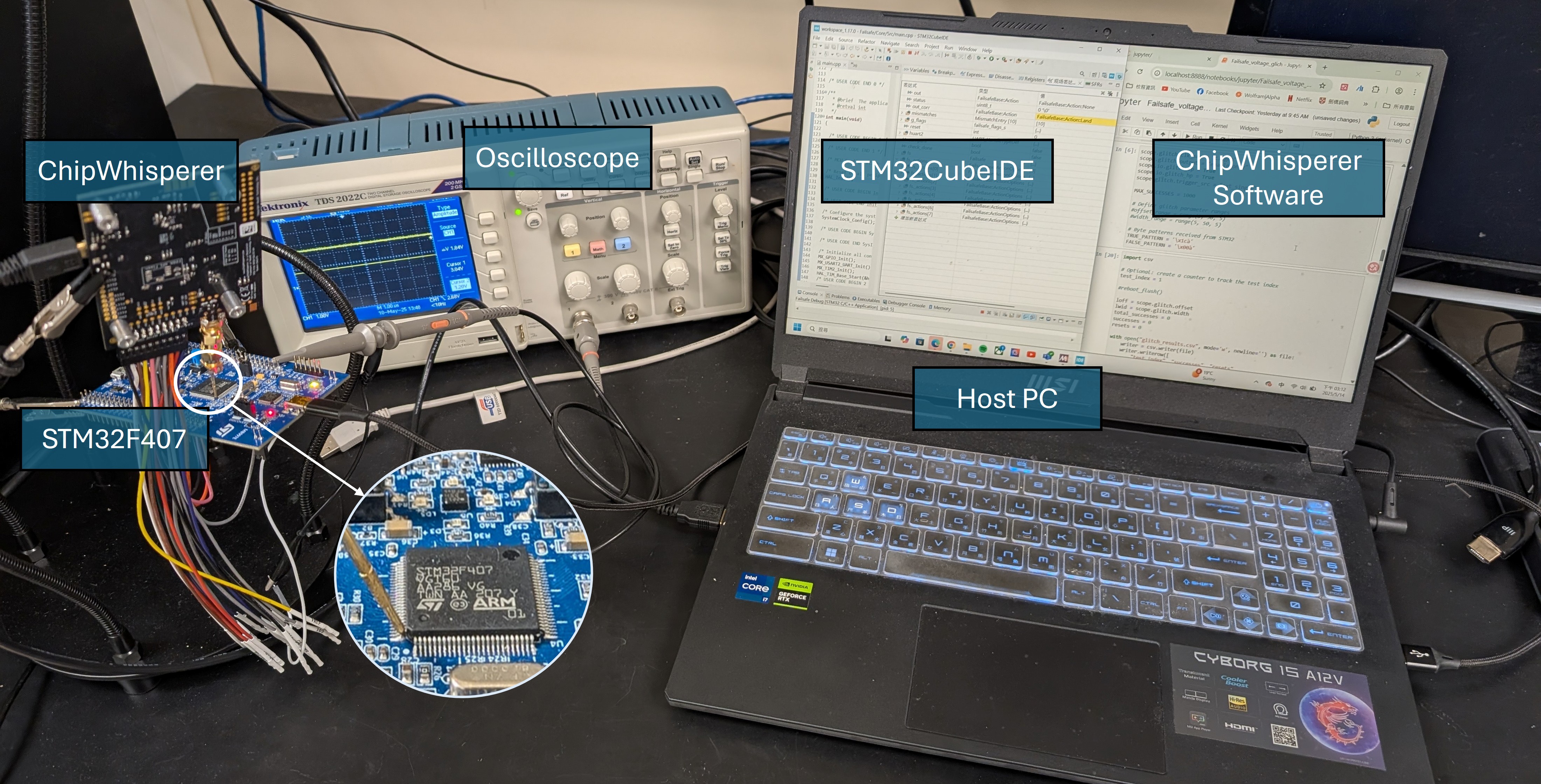}
        \caption{Physical setup for hardware fault injection experiments.}
     
    \caption{Experimental hardware setup}
    \label{fig:hardware_setup}
\end{figure}

The experimental setup consisted of two core components: the ChipWhisperer-Lite fault injection tool and the STM32F407-Discovery target board. The physical architecture is shown in Figure~\ref{fig:hardware_setup}.

\begin{figure}[htbp]
    \centering
    \includegraphics[width=0.5\textwidth]{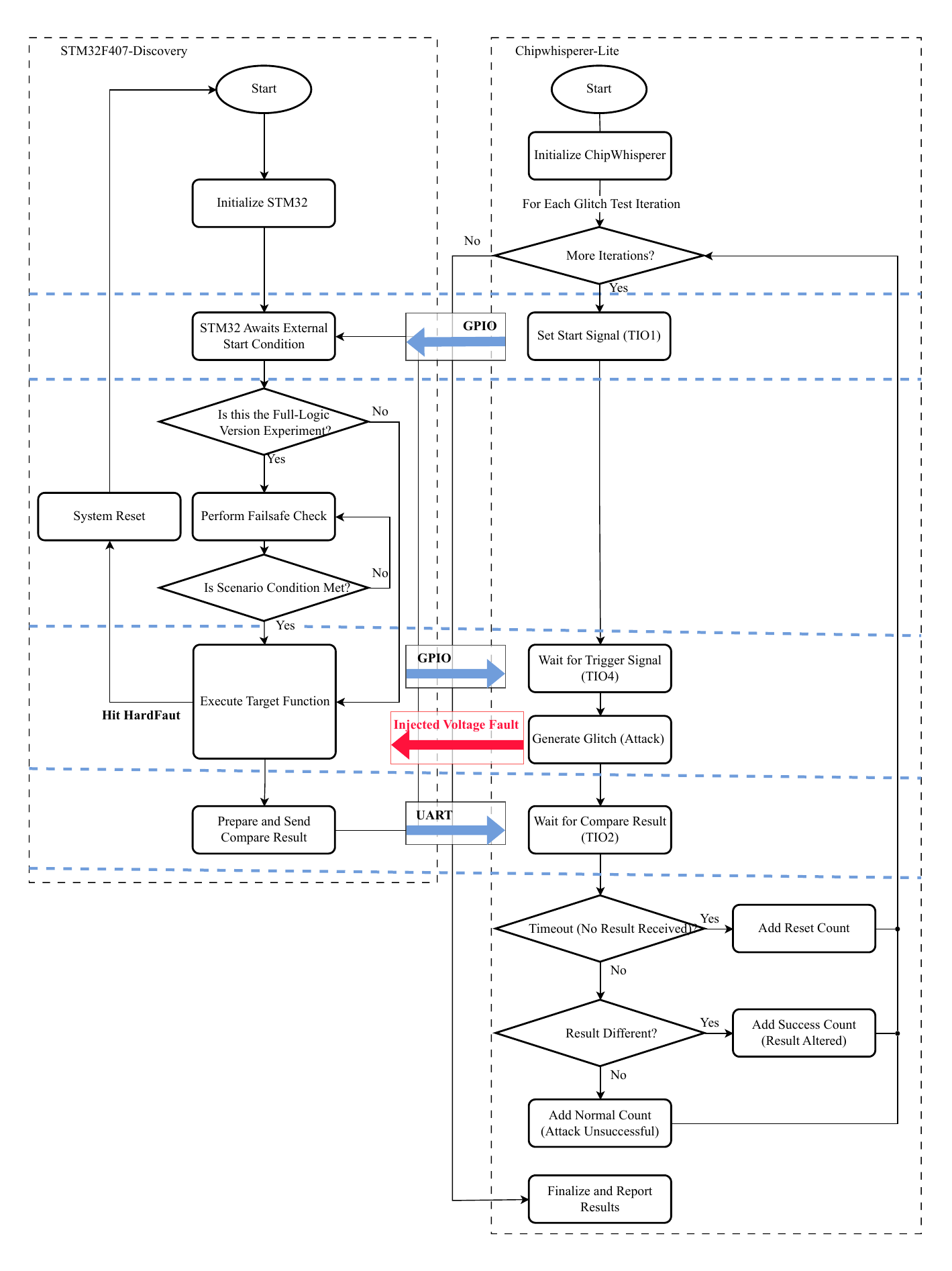}
    \caption{Workflow illustrating the coordinated fault injection process between STM32F407-Discovery and ChipWhisperer-Lite.}
    \label{fig:firmware_flowchart}
\end{figure}

To guarantee precise, reproducible fault injection, critical connections for synchronization and data transmission were established. This implementation followed a strict two-stage synchronization workflow, detailed in Figure~\ref{fig:firmware_flowchart}.

\begin{itemize}
\item Stage 1 – Execution Initiation: 
To initiate each test iteration, the \texttt{TIO1} output pin on the ChipWhisperer-Lite was connected to \texttt{PD14} pin, configured as a general-purpose input, on the STM32F407. This allowed the external system to control the start of code execution.
\item Stage 2 – Glitch Trigger: 
During program execution, the STM32F407 generated a rising-edge trigger signal on its \texttt{PD12} output pin, which was routed to the ChipWhisperer-Lite’s input pin \texttt{TIO4}. This trigger signal served as a reference point for the external offset counter, ensuring precise temporal alignment of the fault injection pulse.
\item UART data transfer: 
To transmit the results of the injection attempt, the \texttt{PA2} transmit pin (USART2) on the STM32F407 was connected to the \texttt{TIO2} receive pin on the ChipWhisperer-Lite. This enabled the transfer of the target's fault response data back to the host computer for downstream analysis.
\end{itemize}

Finally, an oscilloscope was probed directly to an alternative $V_{DD}$ pin on the STM32F407 microcontroller to monitor real-time voltage behavior and visually verify the amplitude and duration of the injected fault pulses against the capacitor network.

%% file: sections/evaluation.tex
\section{Evaluation} 

To systematically evaluate the impact of the voltage glitch, we analyze the results across three distinct operational scenarios within the flight controller's failsafe module: \textbf{RC Signal Loss}, \textbf{Critical Battery Low}, and \textbf{Emergency Battery Low}. The evaluation compares theoretical vulnerabilities identified via software simulation (ARMORY) against empirical data gathered from physical hardware injection (STM32 and ChipWhisperer).

\subsection{Experimental Alignment}
A critical component of this evaluation is aligning the temporal execution of the simulation with the physical hardware. The STM32 microcontroller operates at 168 MHz, whereas the ChipWhisperer platform was configured to 32 MHz. Because ARMORY accurately simulates the STM32 environment, an offset of one clock cycle on the ChipWhisperer corresponds to approximately 5.25 execution cycles within the ARMORY simulation.  

Furthermore, the defined starting points in the ARMORY simulation and the hardware experiment do not perfectly align. The transmission delay between the STM32 and the ChipWhisperer must be accounted for. For isolated function tests (the switch-case logic in RC Signal Loss and Critical Battery) and integrated tests (the Emergency Battery scenario), several instructions execute between the STM32 generating the trigger signal and the actual arrival at the target switch-case function. The target's $V_{DD}$ line requires finite time to physically decay into the critical "risk range" where faults manifest. Finally, to ensure non-destructive testing, the glitch width is artificially extended, often resulting in voltage instability that outlasts the helper function's execution time. Therefore, the primary objective of this evaluation is to analyze the trend correlation between the software vulnerabilities and the hardware exploitability, rather than seeking precise cycle-to-cycle replication. Acknowledging this physical constraint also justifies our progression from analyzing isolated helper functions to testing the fully integrated failsafe module (Scenario 1) to observe how these extended, unpredictable hardware faults propagate in a realistic context.

\subsection{Scenario 1: RC Signal Loss}
This scenario simulates a complete loss of communication between the UAV and the operator. The PX4 firmware evaluates this state using a dedicated failsafe helper function, typically triggering a Return-to-Launch (RTL) protocol. For this evaluation, an injected fault was classified as \textbf{"Successful"} if the observed output parameters (\textit{ActionOptions} : \textit{Action}, \textit{Cause}, \textit{AllowUserTakeover}, \textit{ClearCondition}) deviated from the expected outcome under fault-free execution.

\subsubsection{Software Simulation Findings (ARMORY)}
\begin{figure}[ht]
    \centering
    \includegraphics[width=\linewidth]{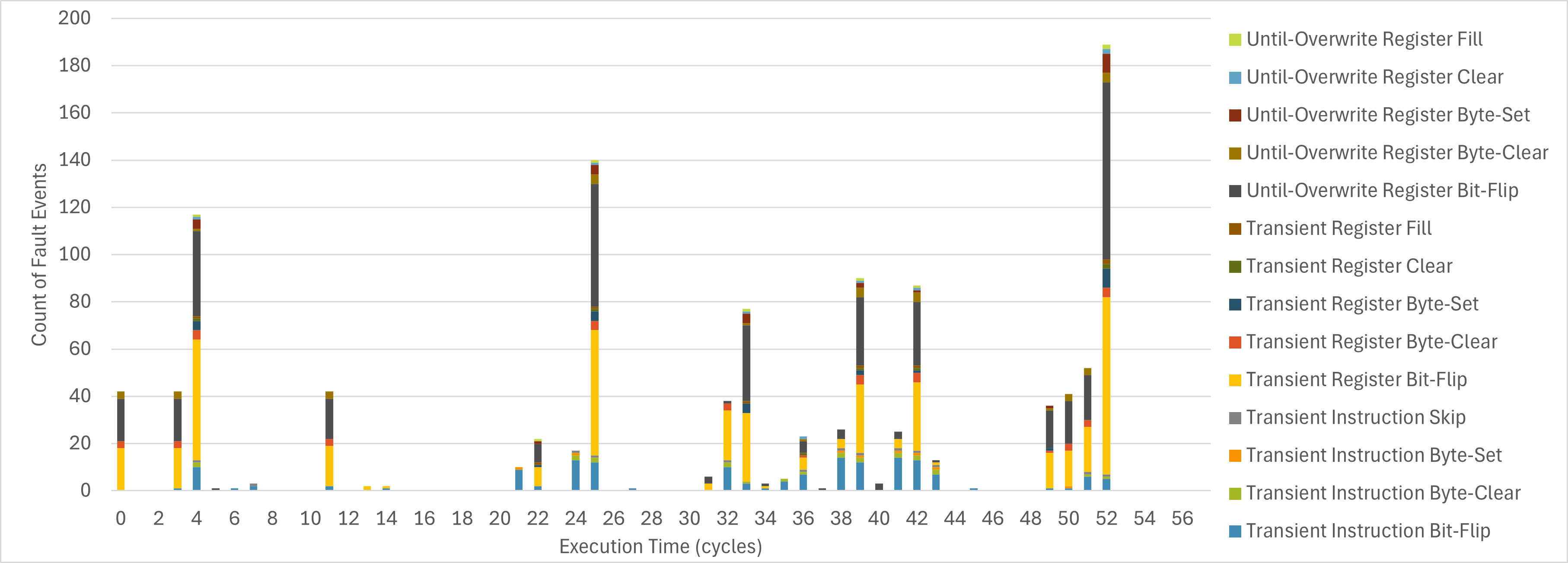}
    \caption{Temporal distribution of successful faults in RC Signal Loss scenario}
    \label{fig:rc_loss_temporal}
\end{figure}

As illustrated in Figure~\ref{fig:rc_loss_temporal}, successful faults demonstrated strong temporal clustering during execution. Prominent peaks emerged between clock cycles 3–4, 24–25, and 50–52. These peaks are largely driven by register-level corruption, specifically Transient Register Bit-Flip and Until-Overwrite Register Bit-Flip fault models.

\subsubsection{Hardware Validation}
The hardware experiment (Figure 5.2) validated the existence of these critical timing windows. Hardware \textit{Offsets} of 4 and 5 cycles consistently exhibited elevated fault success rates across multiple glitch widths. Applying the clock ratio, these physical offsets align well with the vulnerability cluster observed around simulation cycles 21–26. Additionally, an \textit{Offset} of 2 cycles (aligning with ARMORY cycles 11) required a longer glitch width to yield higher \textit{Success} outcomes (blue bars). The data indicates that longer glitches more reliably corrupt critical instruction execution, increasing the likelihood of both HardFault-triggered resets (orange bars) and exploitable faults. \textit{Offset} 0 was excluded from the main dataset because early testing revealed it resulted in a 100\% reset rate, frequently causing the STM32 to suspend unexpectedly.

\begin{figure}[h]
    \centering
    \begin{minipage}{0.32\linewidth}
        \centering
        \includegraphics[width=\linewidth]{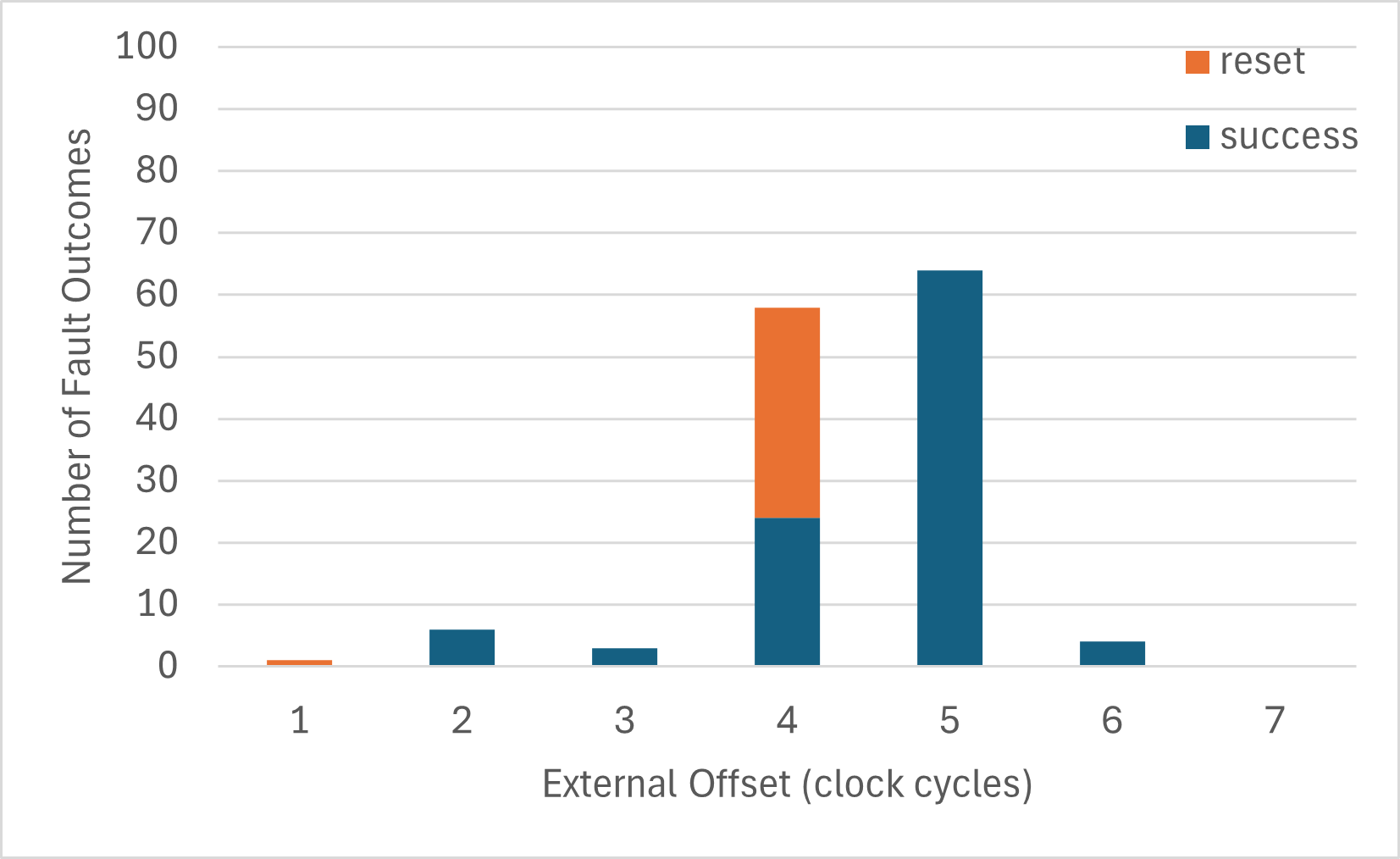}
        \caption*{(a) 1.66\,$\mu$s}
    \end{minipage}
    \hfill
    \begin{minipage}{0.32\linewidth}
        \centering
        \includegraphics[width=\linewidth]{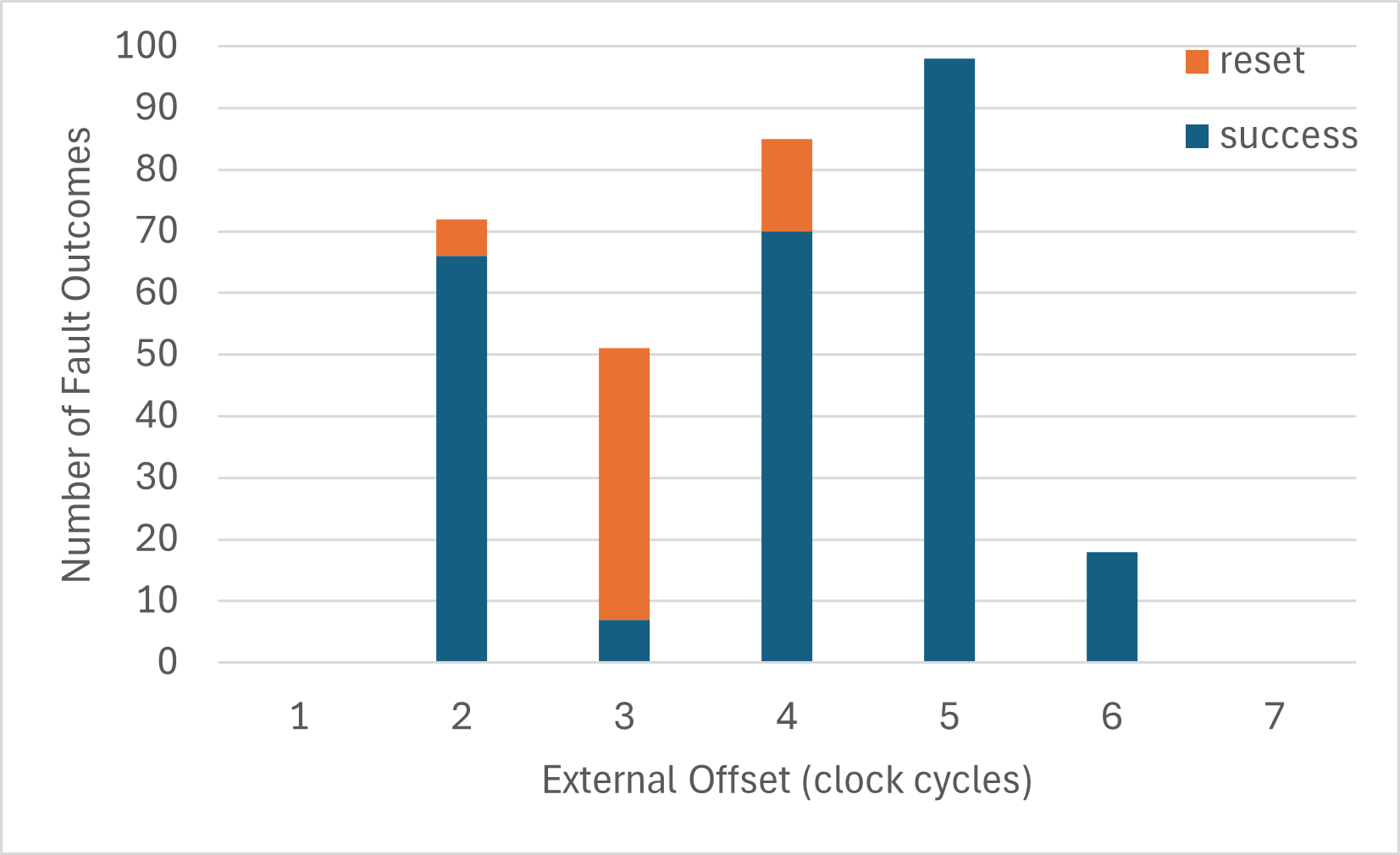}
        \caption*{(b) 1.69\,$\mu$s}
    \end{minipage}
    \hfill
    \begin{minipage}{0.32\linewidth}
        \centering
        \includegraphics[width=\linewidth]{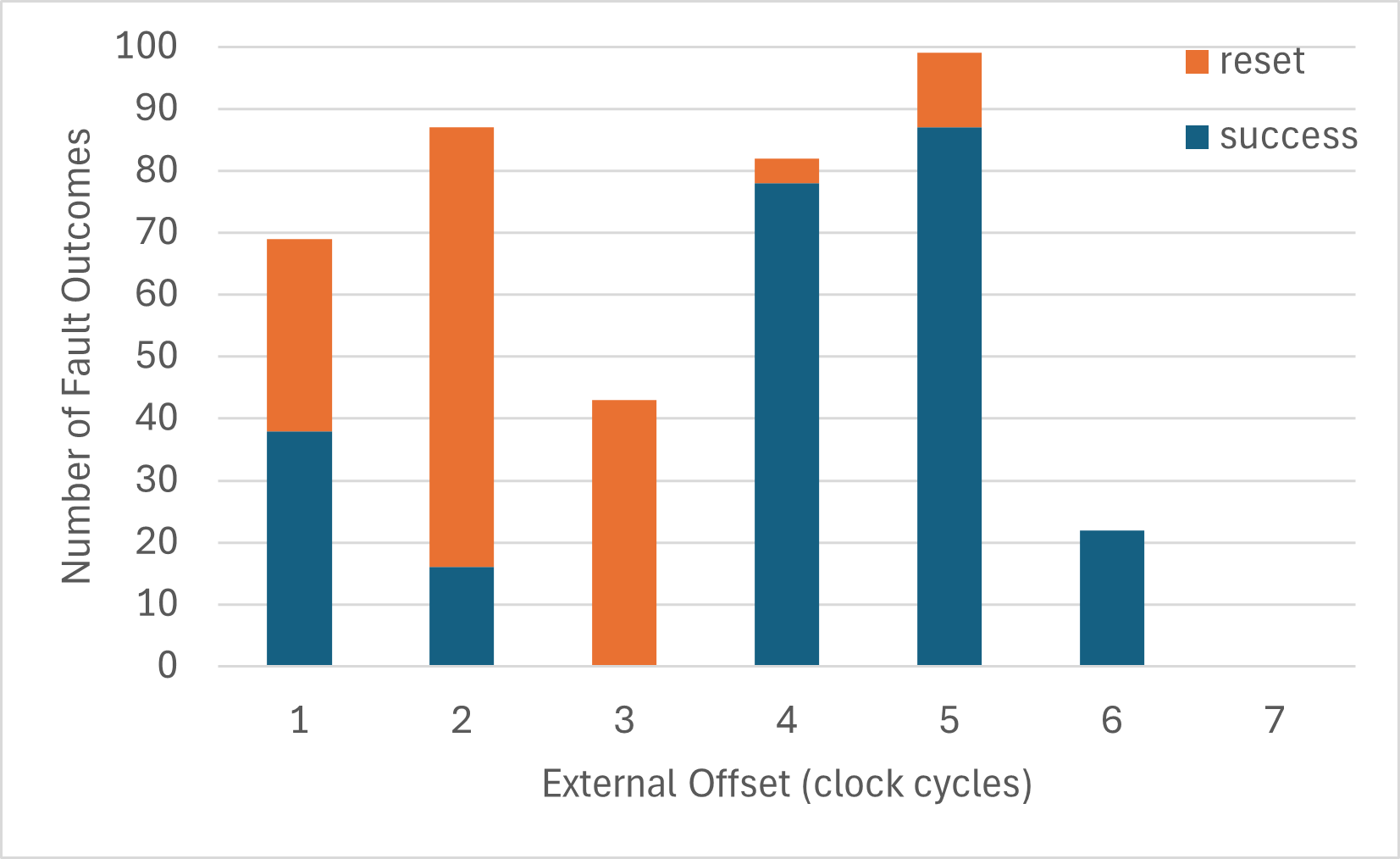}
        \caption*{(c) 1.72\,$\mu$s}
    \end{minipage}
    \caption{Fault injection outcomesin RC Signal Loss scenario}
    \label{fig:rcloss_simplified_glitchsweep}
\end{figure}

\subsubsection{Fault Effect Correlation}
\begin{figure}[ht]
    \centering
    \includegraphics[width=\linewidth]{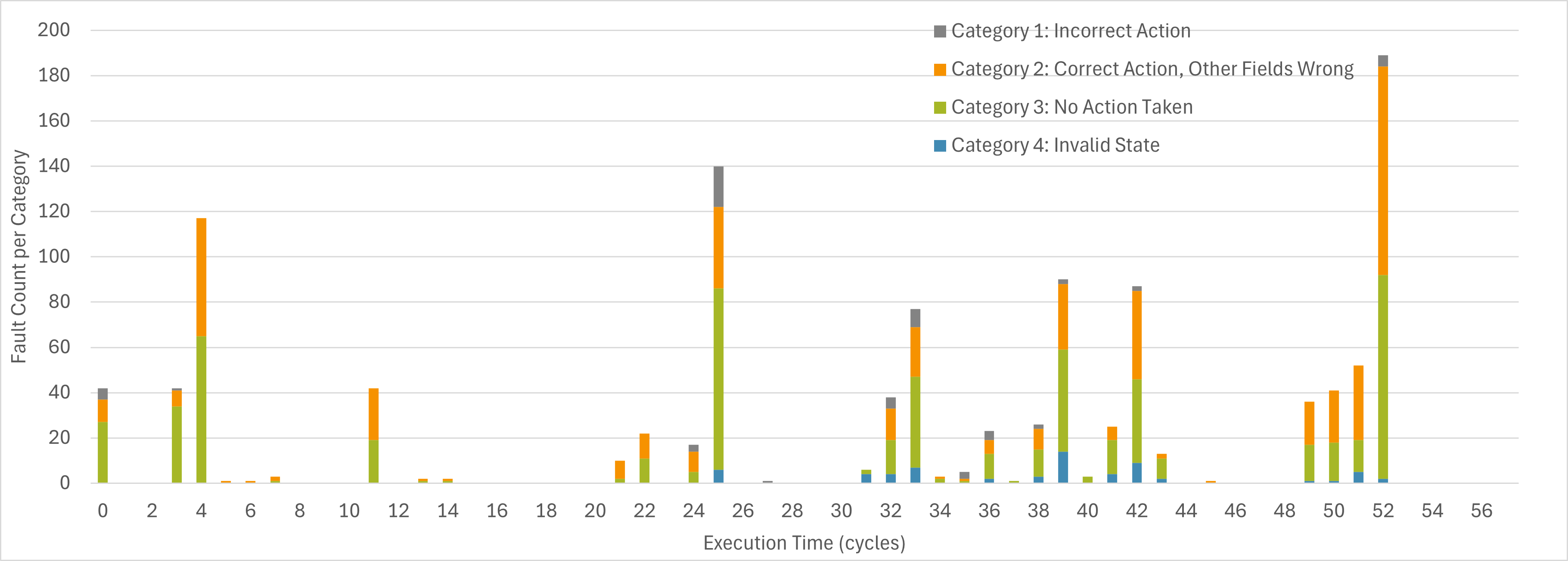}
    \caption{Temporal distribution of fault effects categorized by behavior outcome for RC Signal Loss scenario}
    \label{fig:rc_loss_action}
\end{figure}

\begin{figure}[htbp]
    \centering
    \begin{minipage}{\linewidth}
        \centering
        \includegraphics[width=\linewidth]{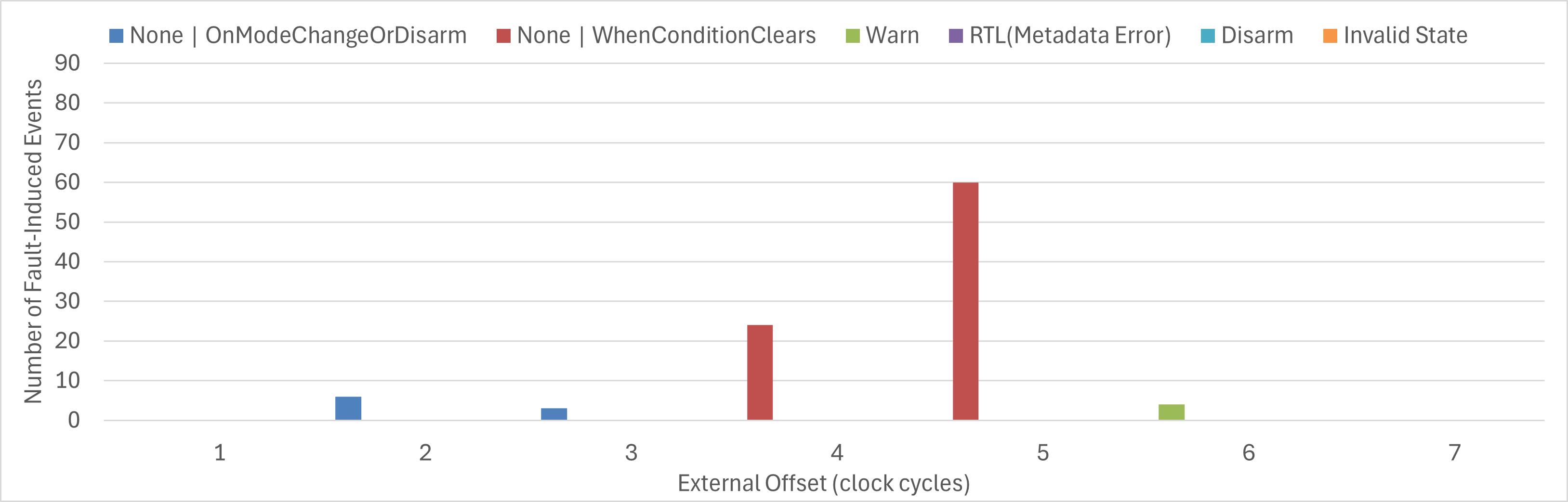}
        \caption*{(a) 1.66\,$\mu$s}
    \end{minipage}
    \hfill
    \begin{minipage}{\linewidth}
        \centering
        \includegraphics[width=\linewidth]{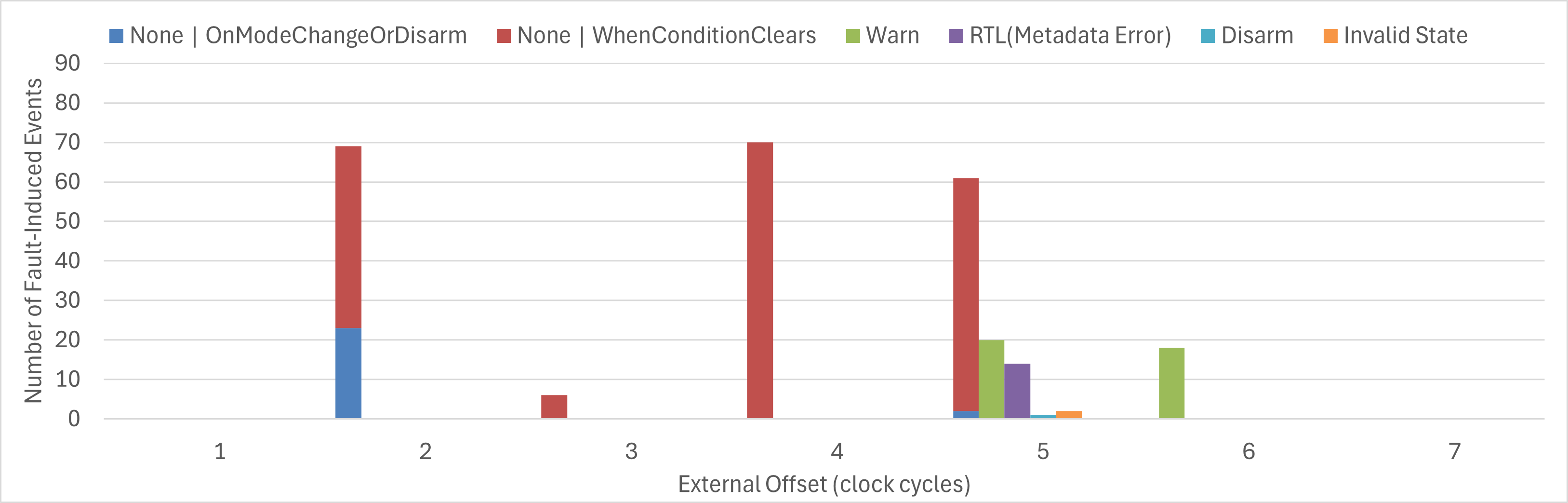}
        \caption*{(b) 1.69\,$\mu$s}
    \end{minipage}
    \hfill
    \begin{minipage}{\linewidth}
        \centering
        \includegraphics[width=\linewidth]{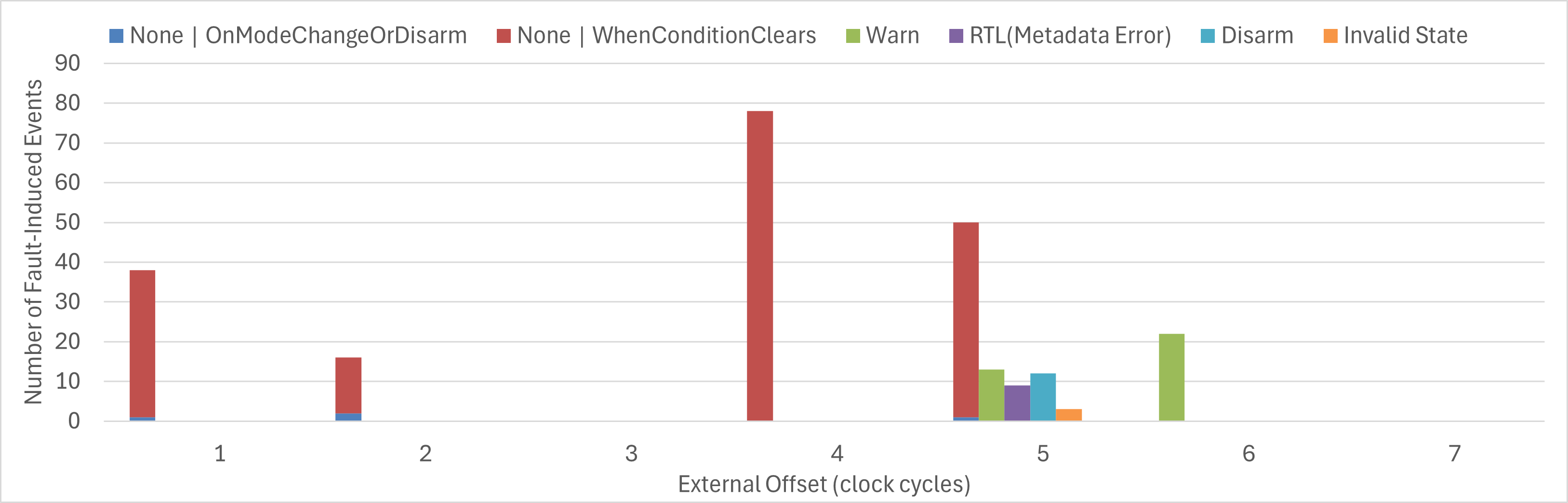}
        \caption*{(c) 1.72\,$\mu$s}
    \end{minipage}
    \caption{Distribution of fault-induced actions in RC Signal Loss scenario under different glitch durations.}
    \label{fig:rcloss_simplified_actions}
\end{figure}

By correlating the simulated data (Figure~\ref{fig:rc_loss_action}) with the observed hardware behaviors (Figure~\ref{fig:rcloss_simplified_actions}), we can map specific fault outcomes directly to their corresponding execution cycles. Our analysis focuses on simulation cycle 11 and cycles 21–25, as these closely align with the temporal windows where faults were successfully injected in the hardware experiment. In the simulation results, both "No Action" and "Correct Action with other error fields" are prominent, with "No Action" occurring at a slightly higher frequency. In the hardware experiment; however, the action analysis demonstrated that the successful exploit was almost exclusively a complete bypass ("No Action") at earlier physical offsets. At a physical offset of 5 cycles (corresponding to cycle 25 in ARMORY) and subjected to longer glitch widths, the hardware exhibited a peak in "No Action" outcomes alongside instances of "Correct Action with other error fields" (RTL with metadata error), "Error Action," and "Invalid State." This directly mirrors the simulation data, as cycle 25 in ARMORY also observes this exact same diverse distribution of outcomes. 

\subsection{Scenario 2: Battery Low in Critical}
The battery level failsafe is activated when the drone's battery capacity drops below predefined threshold levels. In a "Critical" state, PX4 typically recommends configuring the system to use RTL mode to safely return the drone. Similar to Scenario 1, a fault was considered \textbf{"Successful"} if the observed output behavior deviated from the expected (\textit{ActionOptions} outcome.

\subsubsection{Software Simulation Findings (ARMORY)}
\begin{figure}[ht]
    \centering
    \includegraphics[width=\linewidth]{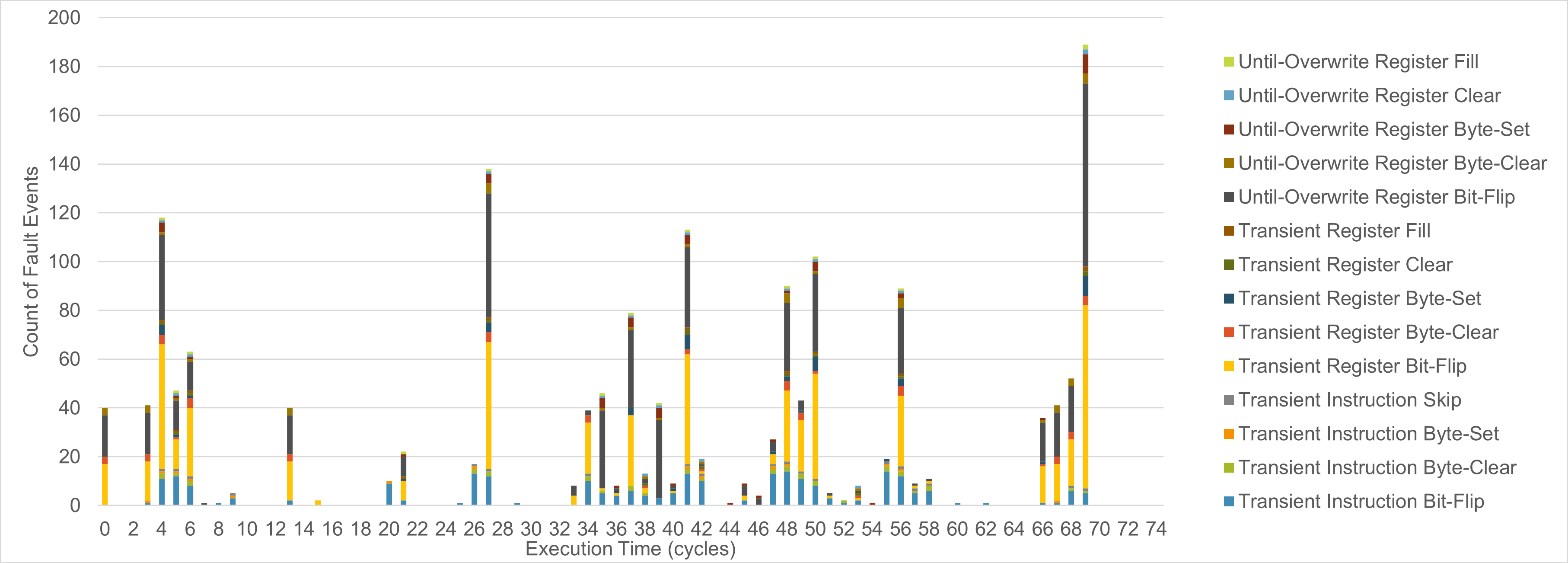}
    \caption{Temporal distribution of successful faults in Battery Critical scenario}
    \label{fig:batt_critical_temporal}
\end{figure}

As shown in Figure~\ref{fig:batt_critical_temporal},  prominent primary vulnerability peaks emerged between cycles 4–6, 26–27, and 66–69. A secondary, continuous density of faults with several minor peaks was observed between cycles 33–58, though significantly less concentrated than the primary three windows.

\subsubsection{Hardware Validation}
\begin{figure}[h]
    \centering
    \begin{subfigure}[b]{0.15\textwidth} 
        \centering
        \includegraphics[width=\textwidth]{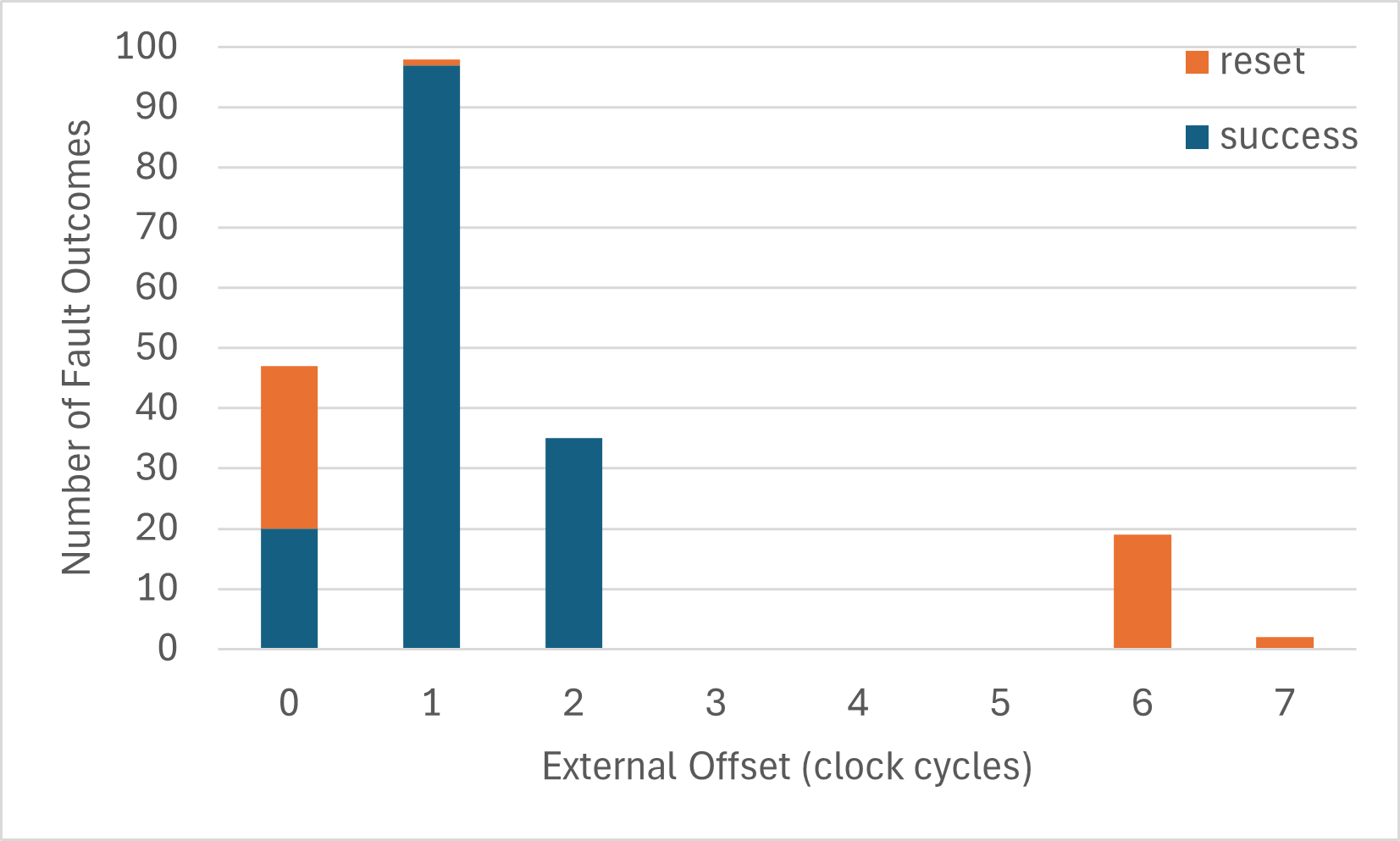} 
        \caption{1.59 µs}
        \label{fig:si_btcr_outcomes_1_59}
    \end{subfigure}
    \hfill
    \begin{subfigure}[b]{0.15\textwidth}
        \centering
        \includegraphics[width=\textwidth]{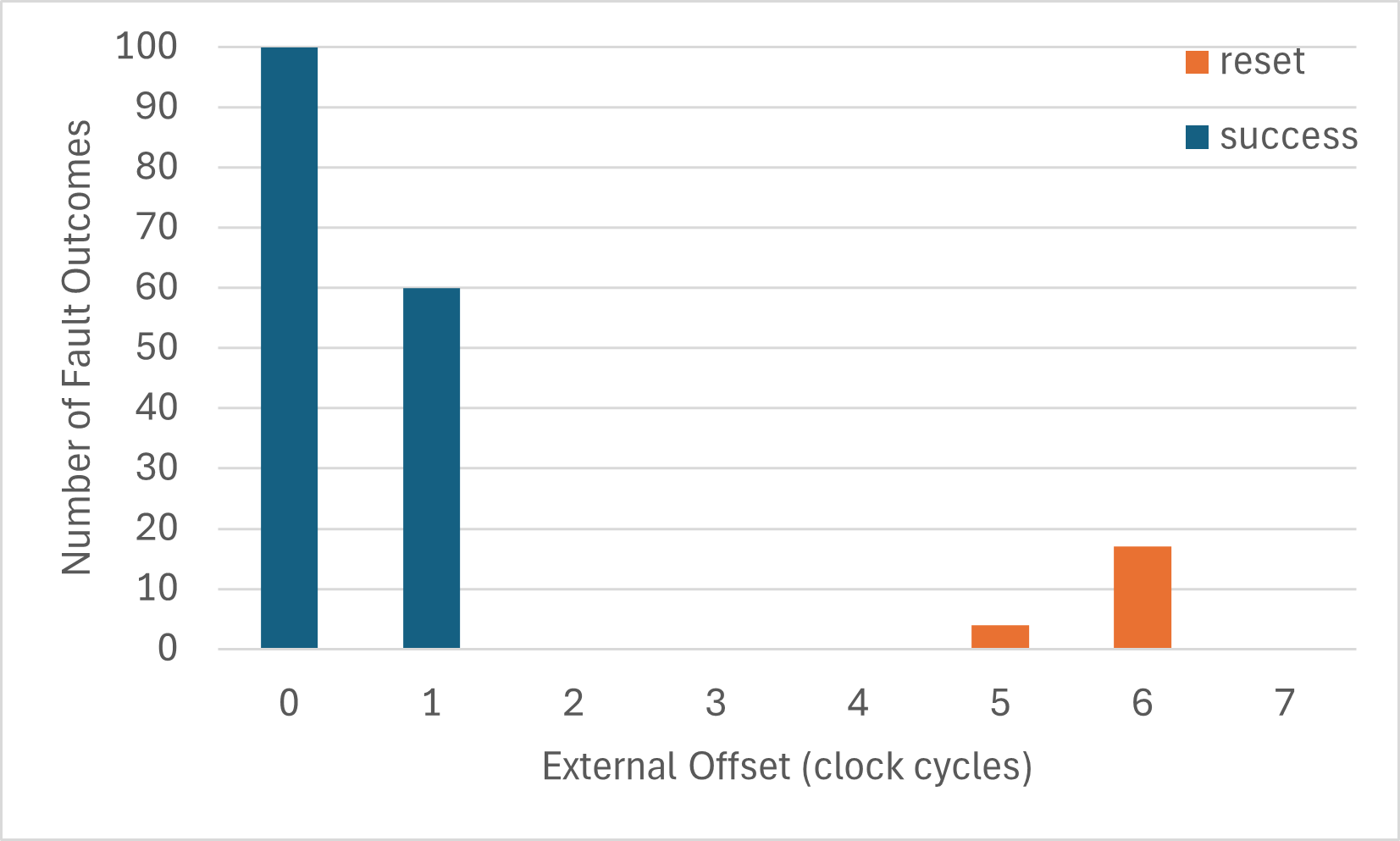} 
        \caption{1.63 µs}
        \label{fig:si_btcr_outcomes_1_63}
    \end{subfigure}
    \hfill
    \begin{subfigure}[b]{0.15\textwidth} 
        \centering
        \includegraphics[width=\textwidth]{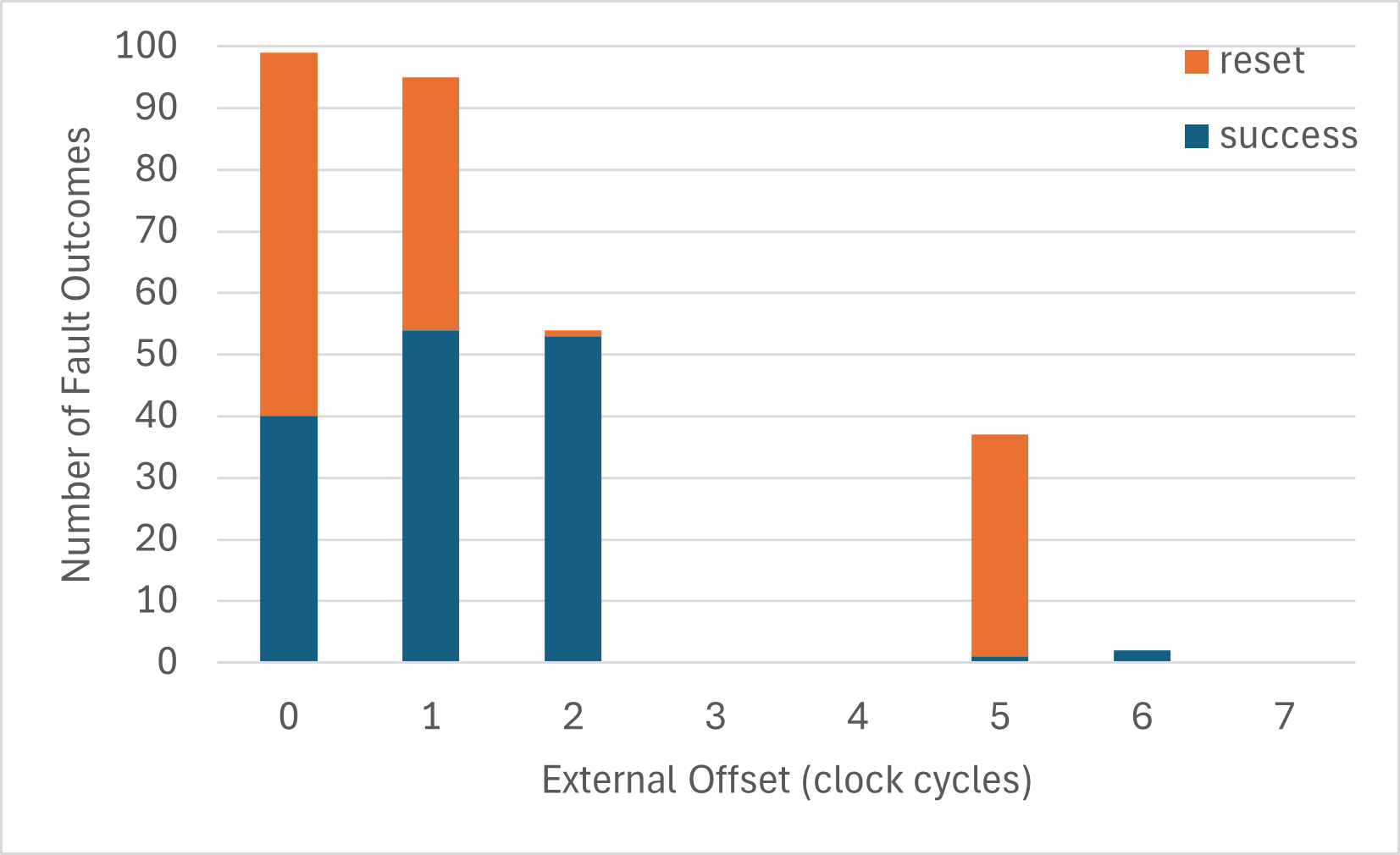} 
        \caption{1.66 µs}
        \label{fig:si_btcr_outcomes_1_66}
    \end{subfigure}

    \vspace{1em} 

    \begin{subfigure}[b]{0.15\textwidth} 
        \centering
        \includegraphics[width=\textwidth]{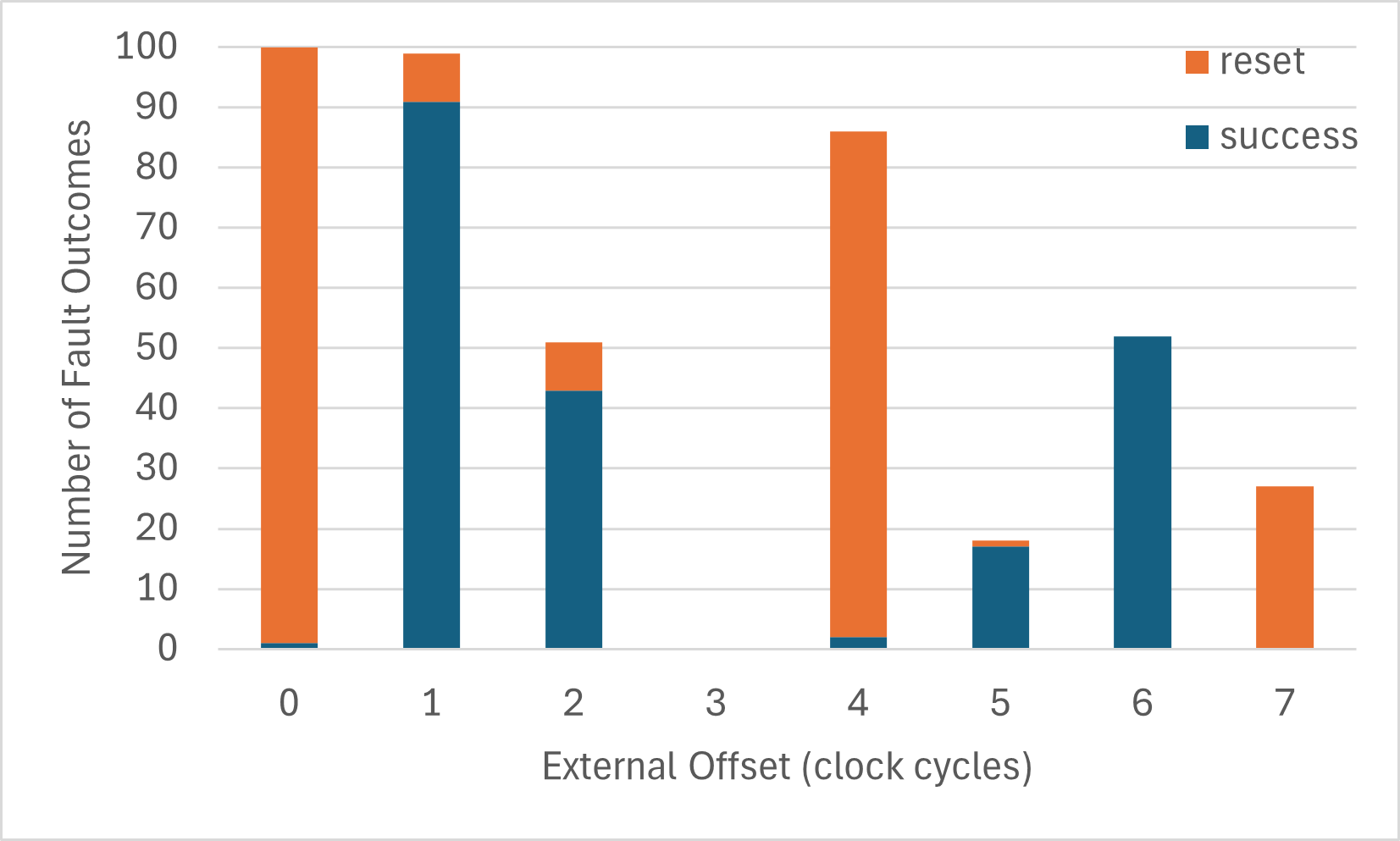}
        \caption{1.69 µs}
        \label{fig:si_btcr_outcomes_1_69}
    \end{subfigure}
    \hfill
    \begin{subfigure}[b]{0.15\textwidth} 
        \centering
        \includegraphics[width=\textwidth]{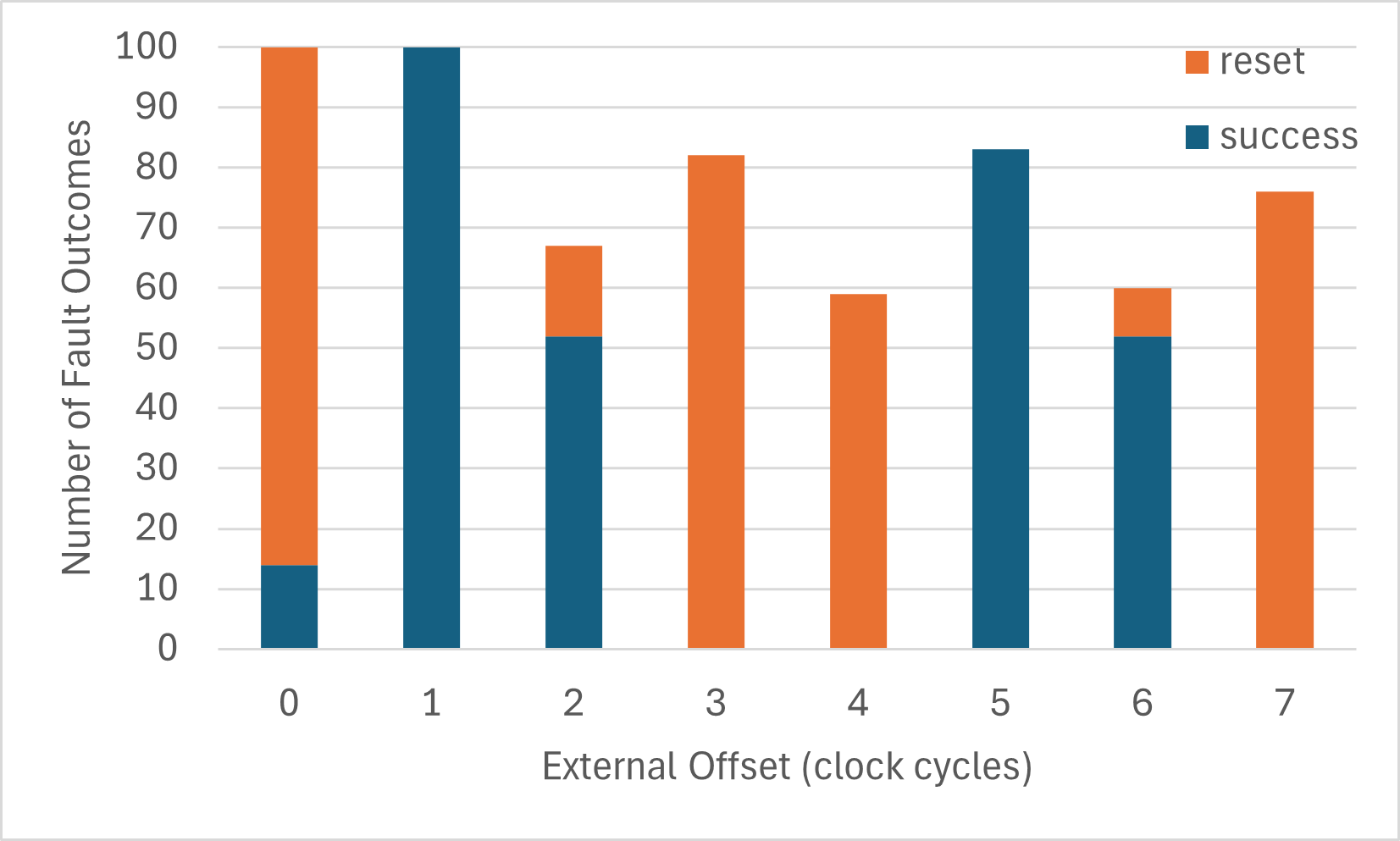}
        \caption{1.72 µs}
        \label{fig:si_btcr_outcomes_1_72}
    \end{subfigure}
    \caption{Distribution of aggregate fault outcomes (Success, Reset) versus \textit{External Offset} for Battery Critical scenario.}
    \label{fig:si_btcr_outcomes_vs_offset}
\end{figure}

The hardware experiment (Figure~\ref{fig:si_btcr_outcomes_vs_offset}) successfully validated the early and late execution vulnerabilities. A physical offset of 1 cycle consistently exhibited elevated fault success rates across multiple glitch widths, aligning closely with the vulnerability cluster observed around simulation cycles 4–6. Interestingly, an \textit{Offset} of 0 cycles yielded a significant success rate at a specific glitch width of 1.63 $\mu$s. Later in the execution, physical offset of 5 cycles (aligning with ARMORY cycles 26–27) yielded high success rates when subjected to longer glitch widths. Consistent with the RC Signal Loss scenario, increasing the glitch width elevated the likelihood of both exploitable faults and HardFault-triggered resets.

\subsubsection{Fault Effect Correlation}
\begin{figure}[ht]
    \centering
    \includegraphics[width=\linewidth]{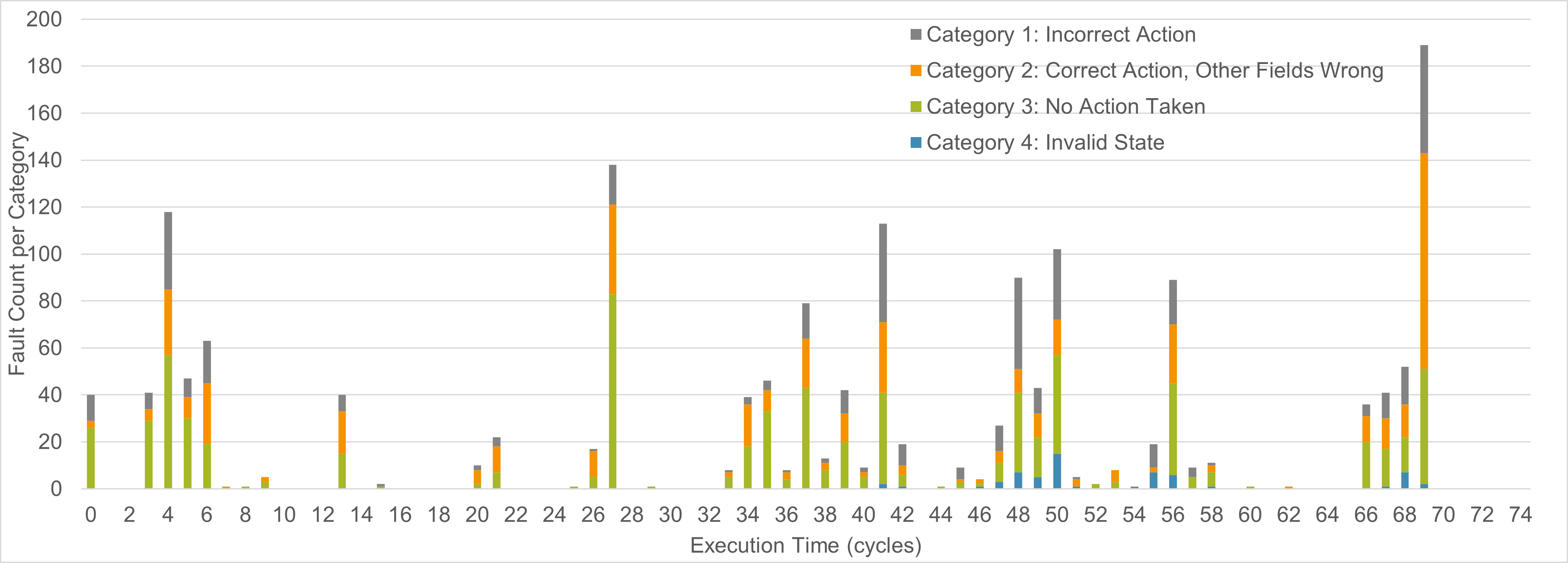}
    \caption{Temporal distribution of fault effects categorized by behavior outcome for Battery Critical scenario}
    \label{fig:batt_critical_action}
\end{figure}

\begin{figure}[htbp]
    \centering
    \begin{minipage}{0.45\linewidth}
        \centering
        \includegraphics[width=\linewidth]{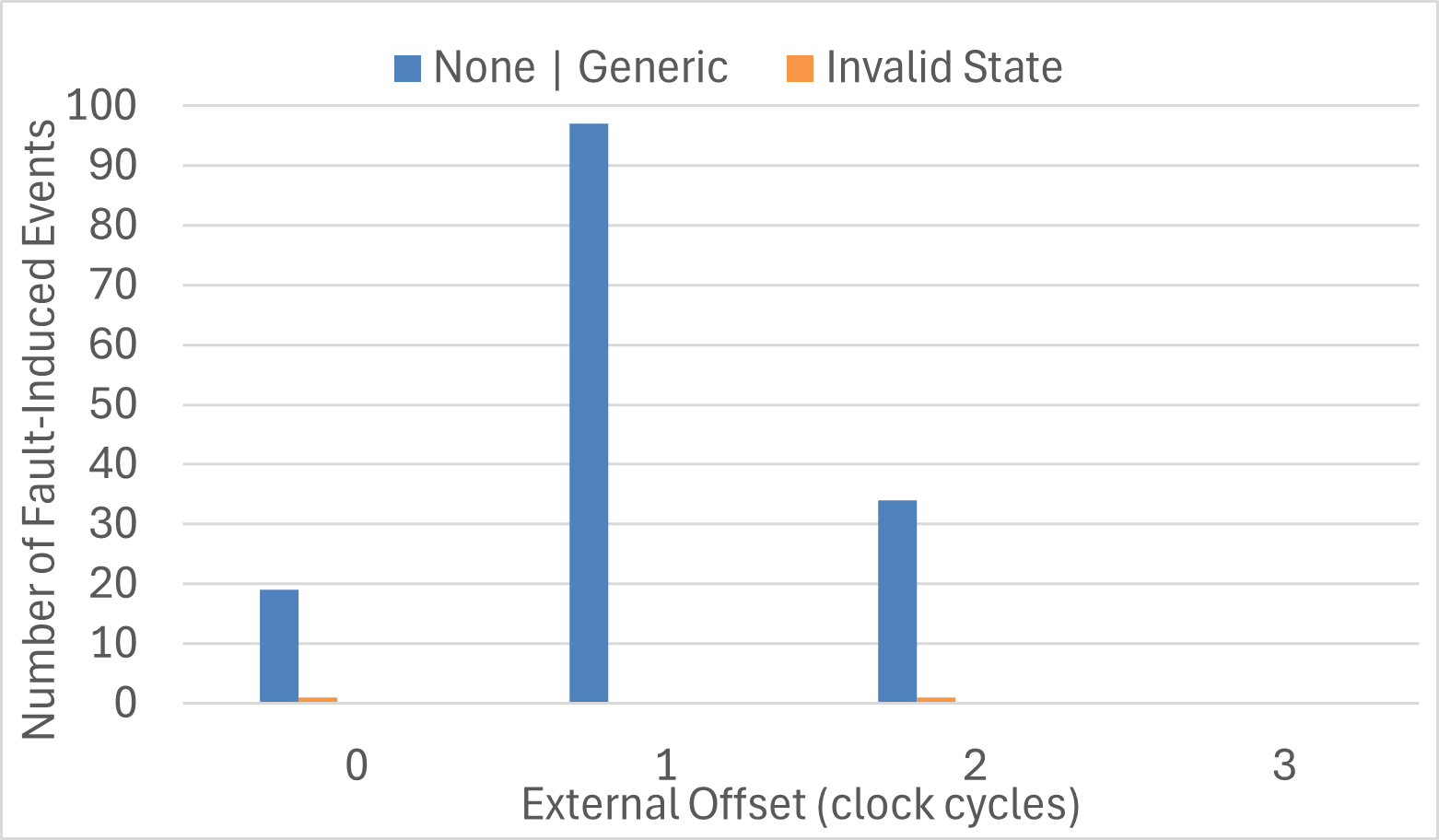}
        \caption*{(a) 1.59\,$\mu$s (Offsets 0-3)}
    \end{minipage}
    \hfill
    \begin{minipage}{0.45\linewidth}
        \centering
        \includegraphics[width=\linewidth]{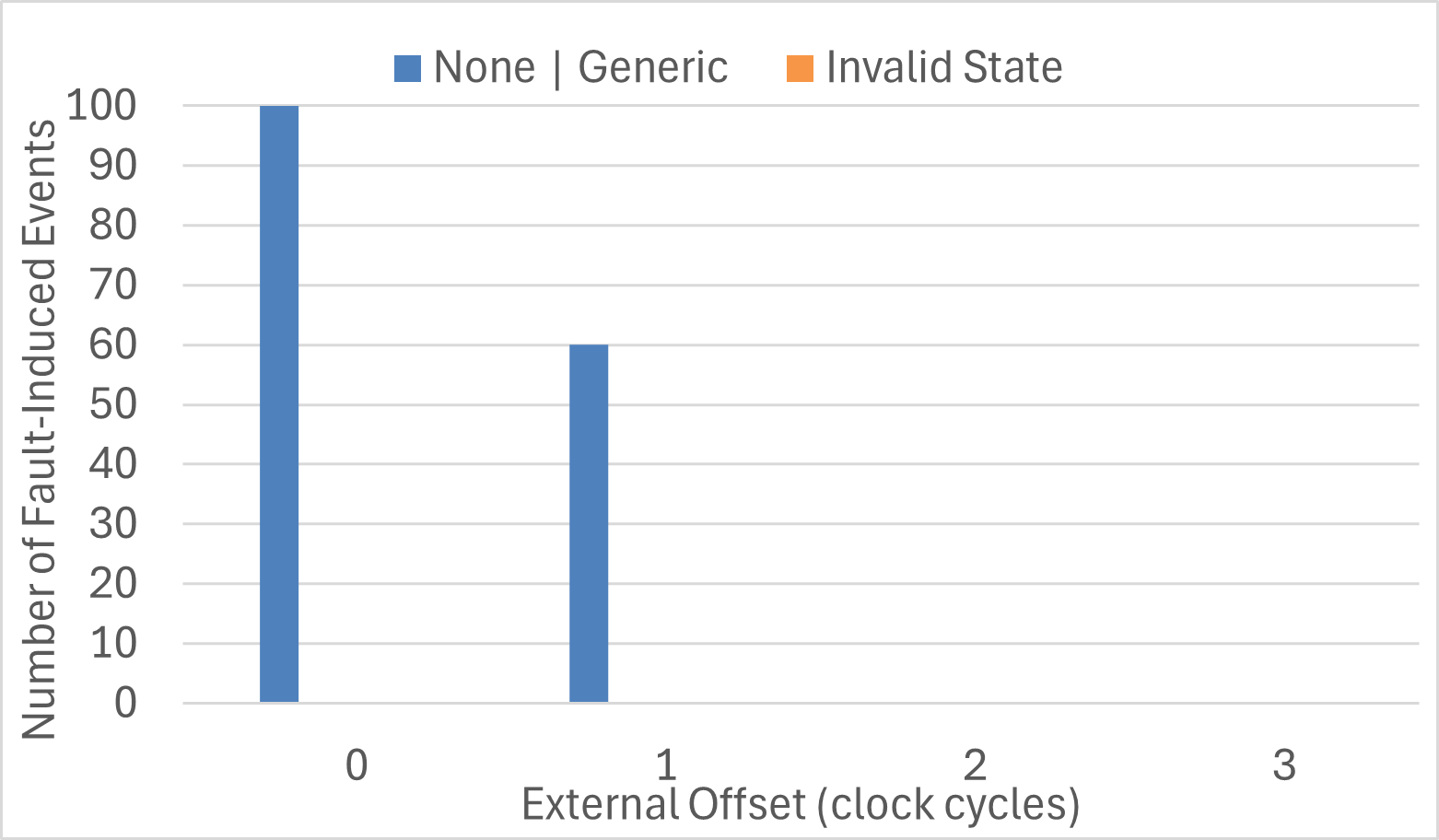}
        \caption*{(b) 1.63\,$\mu$s (Offsets 0-3)}
    \end{minipage}
    \hfill
    \begin{minipage}{0.95\linewidth}
        \centering
        \includegraphics[width=\linewidth]{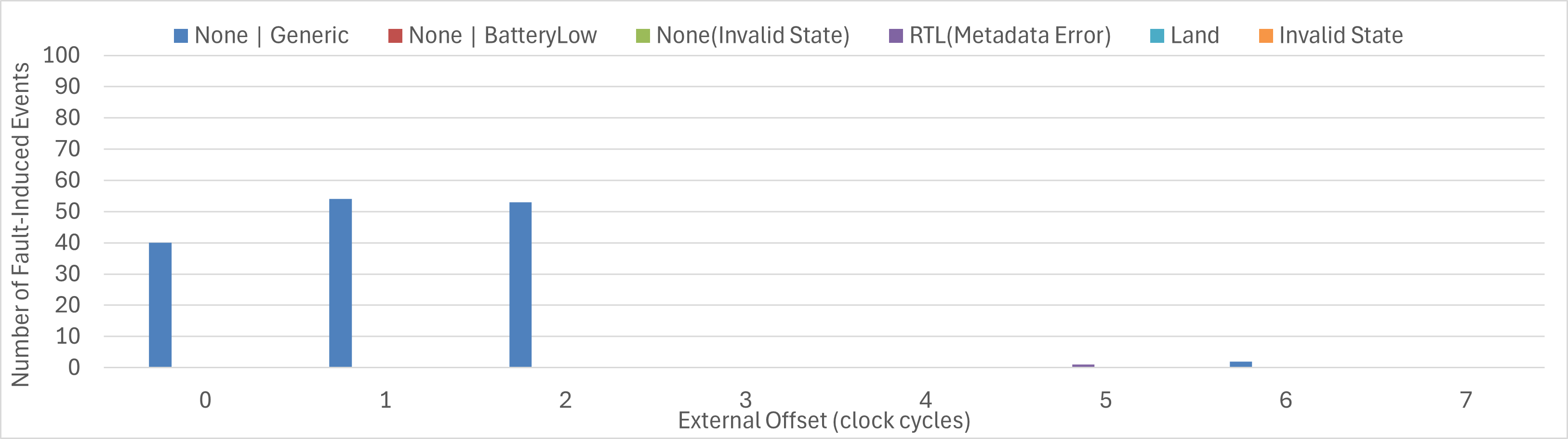}
        \caption*{(c) 1.66\,$\mu$s(Offsets 0-7)}
    \end{minipage}

    \begin{minipage}{0.95\linewidth}
        \centering
        \includegraphics[width=\linewidth]{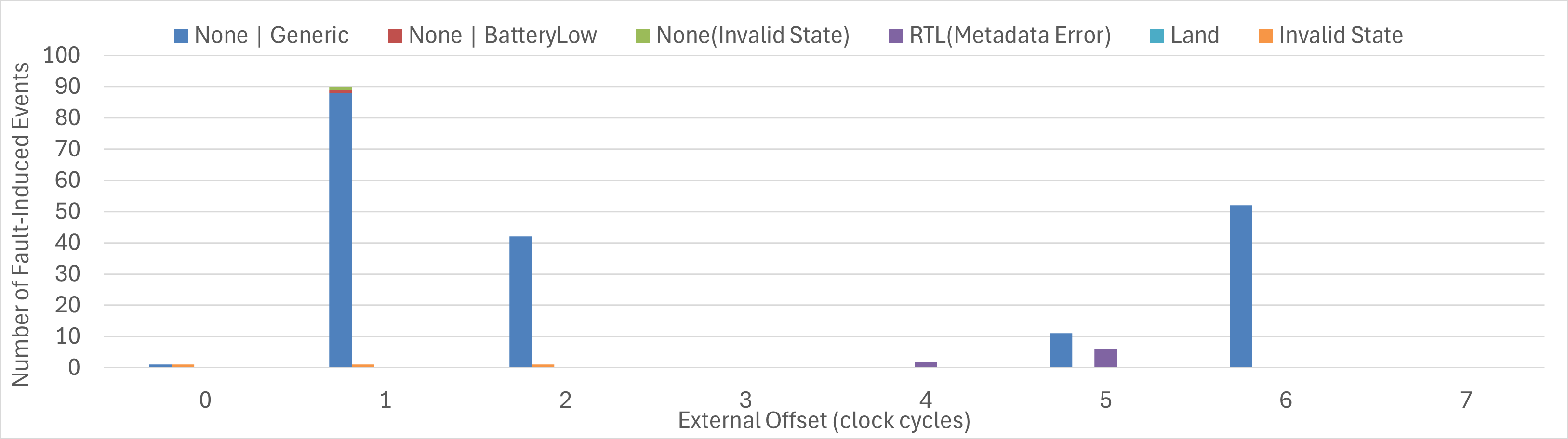}
        \caption*{(d) 1.69\,$\mu$s(Offsets 0-7)}
    \end{minipage}
    \hfill
    \begin{minipage}{0.95\linewidth}
        \centering
        \includegraphics[width=\linewidth]{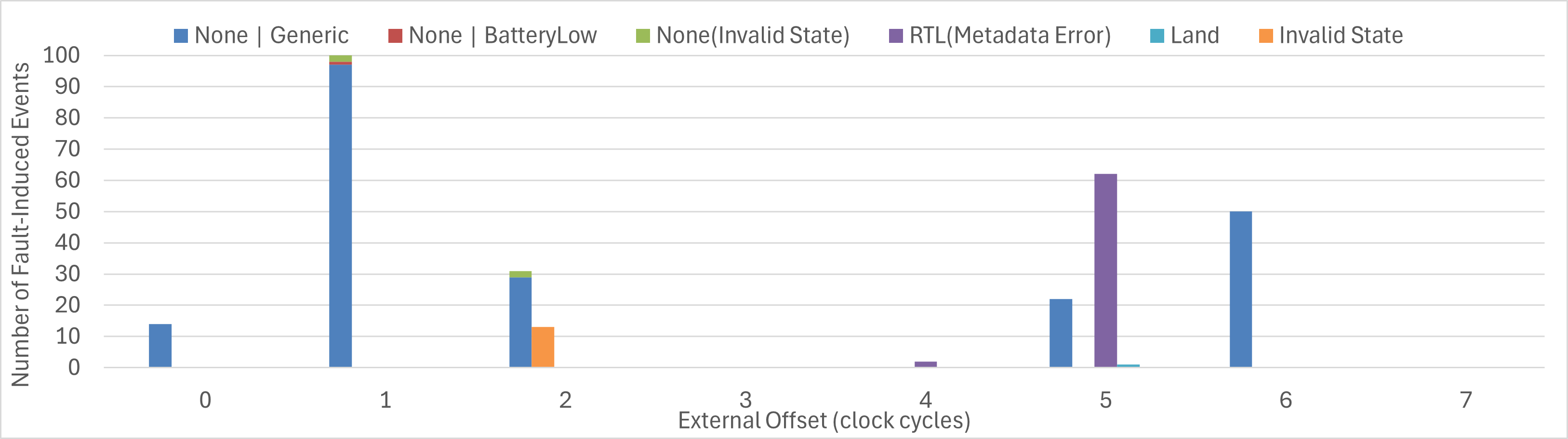}
        \caption*{(e) 1.72\,$\mu$s(Offsets 0-7)}
    \end{minipage}
    
    \caption{Distribution of specific observed Failsafe Action Modes and Outcomes versus \textit{External Offset} for Battery Critical scenario.}
   \label{fig:si_btcr_action_outcomes}
\end{figure}

Mapping the simulated fault effects (Figure~\ref{fig:batt_critical_action}) to the observed physical hardware effects (Figure~\ref{fig:si_btcr_action_outcomes}) reveals nuanced execution behaviors. In the simulation, faults injected at cycles 4–6 resulted in "No Action" at a slightly higher frequency than the other two categories, with zero instances of invalid states. In the hardware experiment (Offset 1), "No Action" remained the most prominent result, but several "Invalid States" were also observed. 
This discrepancy confirms our methodology constraint: because the voltage instability caused by the glitch is long-lasting, the actual instruction fault can unpredictably trigger later in the execution pipeline, resulting in these physical invalid states. For simulation cycles 26–27, the data showed "No Action" as the primary outcome, followed by "Correct Action with other error fields." In the physical experiment (Offset 5); however, it was easier to induce this secondary state, recorded as "RTL (metadata error)" in the hardware findings.

\subsection{Scenario 3: Battery Low in Emergency}
This scenario evaluates the battery level failsafe function operating in conjunction with broader failsafe module code. In an "Emergency" state, PX4 recommends configuring the system to use Land mode, commanding the drone to land immediately. A fault was considered \textbf{"Successful"} if the observed behavior deviated from the expected immediate landing protocol.

\subsubsection{Software Simulation Findings (ARMORY)}
\begin{figure}[ht]
    \centering
    \includegraphics[width=\linewidth]{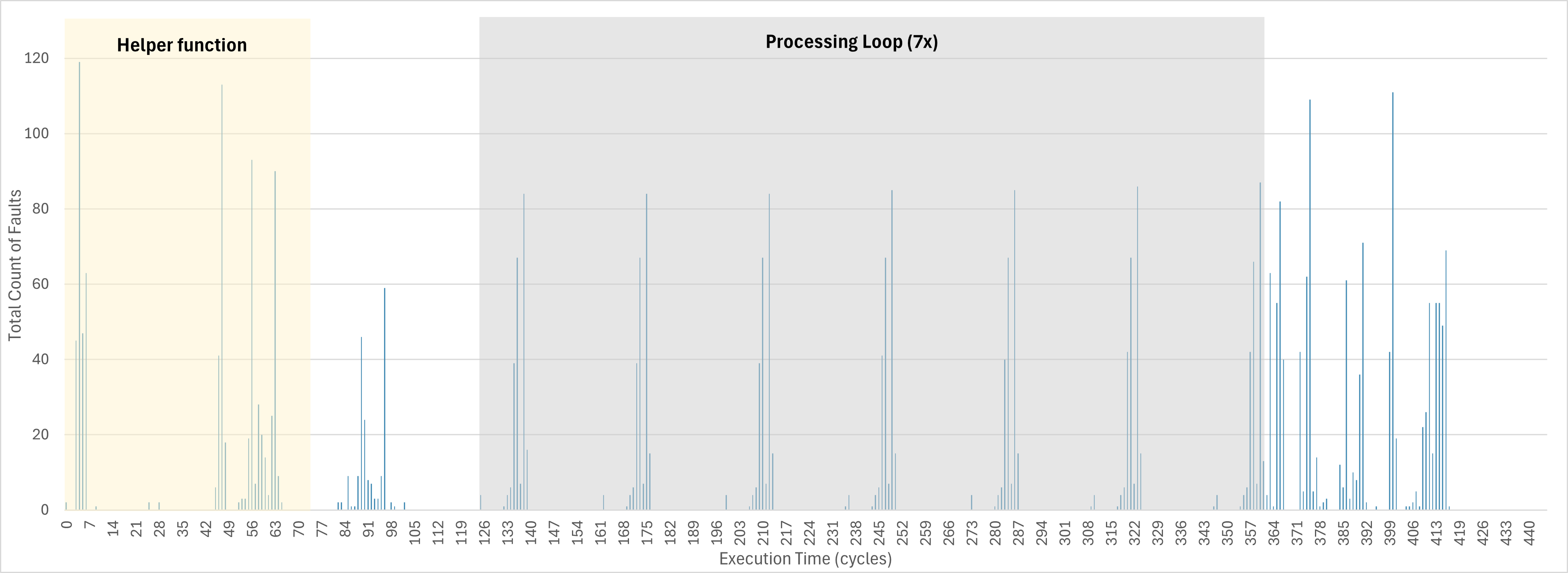}
    \caption{Temporal distribution of successful faults in Battery Emergency scenario}
    \label{fig:batt_emergency_full}
\end{figure}

Unlike previous isolated tests, this simulation (Figure~\ref{fig:batt_emergency_full}) began at the switch-case function label, incorporating a broader execution context. Out of 26,976 injected faults, 1,363 resulted in exploitable outcomes. The most prominent fault activity occurred early in the trace (cycles 0 to 73), corresponding directly to the execution of the battery helper function. Following this, the trace entered a Processing Loop, reflected by a cyclic pattern of fault activity between roughly 130 to 370 clock cycles. 

\begin{figure}[ht]
    \centering
    \includegraphics[width=\linewidth]{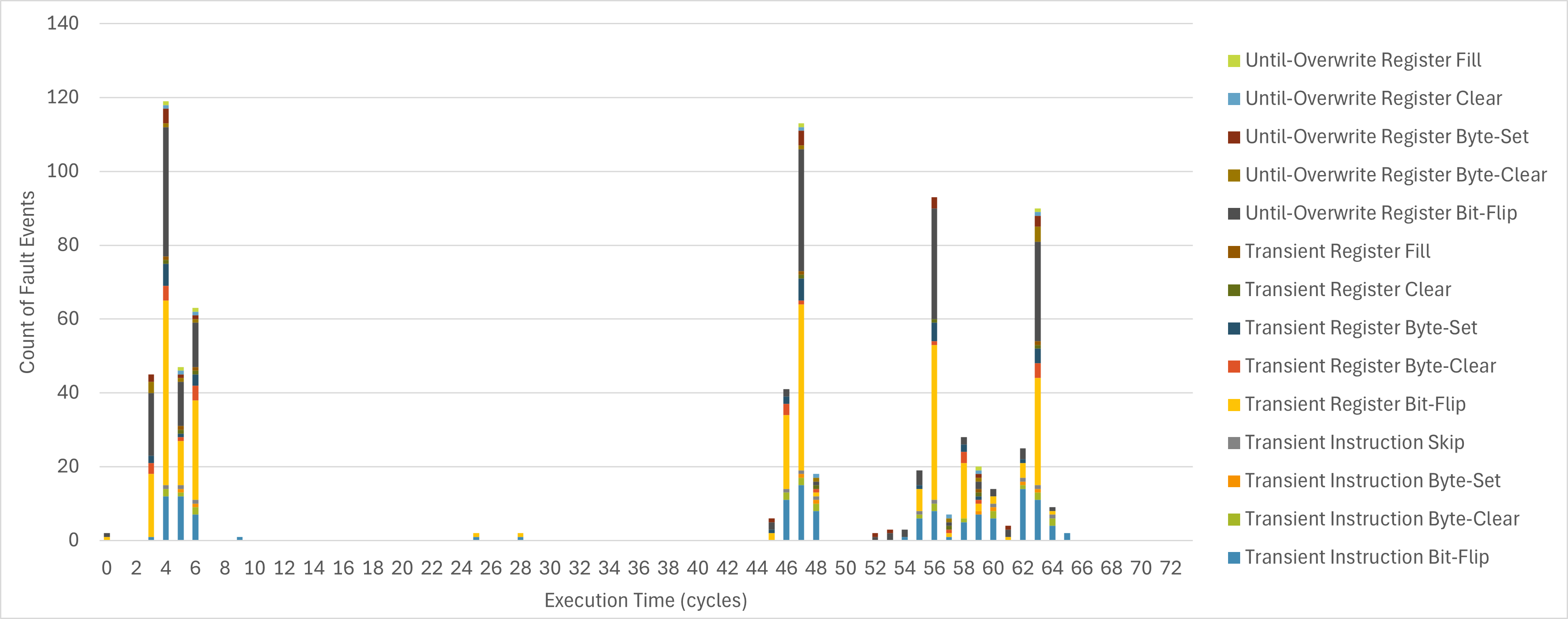}
    \caption{Temporal distribution of successful faults within the helper function execution window in Battery Emergency scenario}
    \label{fig:fs_bt_em_stacked}
\end{figure}

When extracting just the helper function for focused analysis (Figure~\ref{fig:fs_bt_em_stacked}), exploitable outcomes concentrated heavily in the initial clock cycles (3 to 7) and toward the later portion of the window (45 to 64), driven primarily by bit-flip faults targeting registers and instructions.

\subsubsection{Hardware Validation}
\begin{figure}[htbp]
\centering
\begin{subfigure}[b]{0.2\textwidth}
\centering
\includegraphics[width=\textwidth]{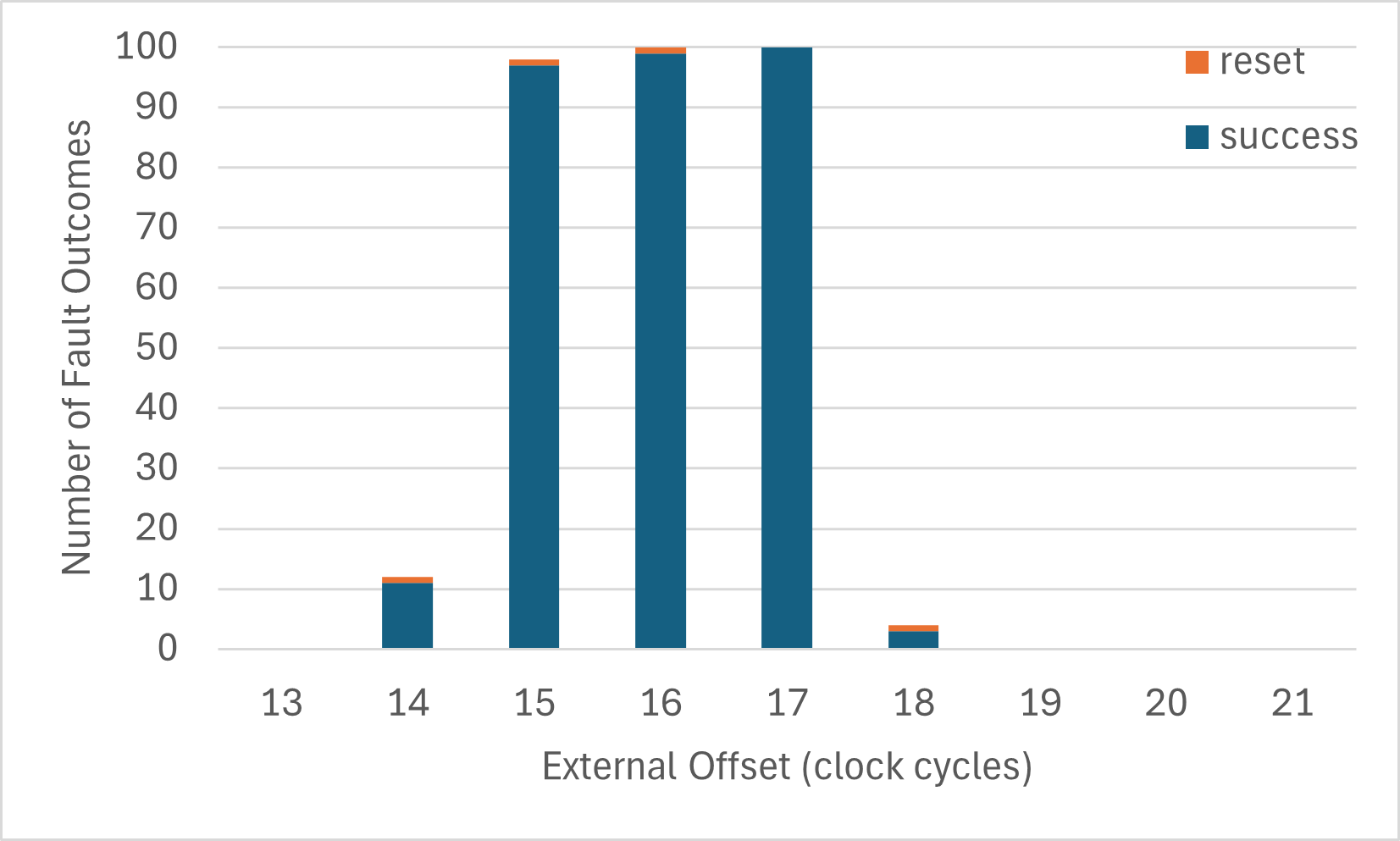} 
\caption{1.69 µs}
\label{fig:btem_outcomes_1_69}
\end{subfigure}
\hfill
\begin{subfigure}[b]{0.2\textwidth} 
\centering
\includegraphics[width=\textwidth]{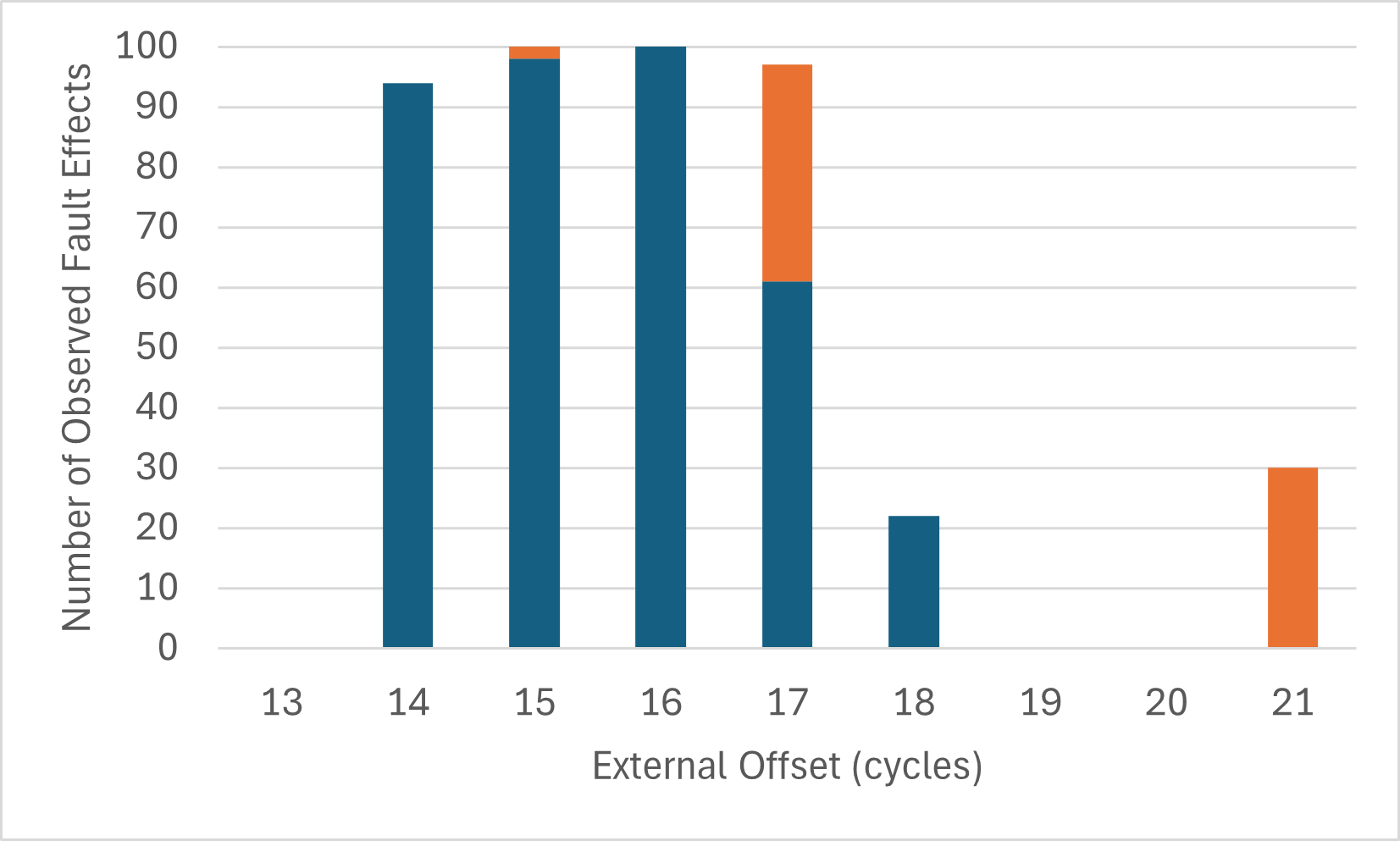}
\caption{1.72 µs}
\label{fig:btem_outcomes_1_72}
\end{subfigure}

\vspace{1em} 

\begin{subfigure}[b]{0.2\textwidth} 
    \centering
    \includegraphics[width=\textwidth]{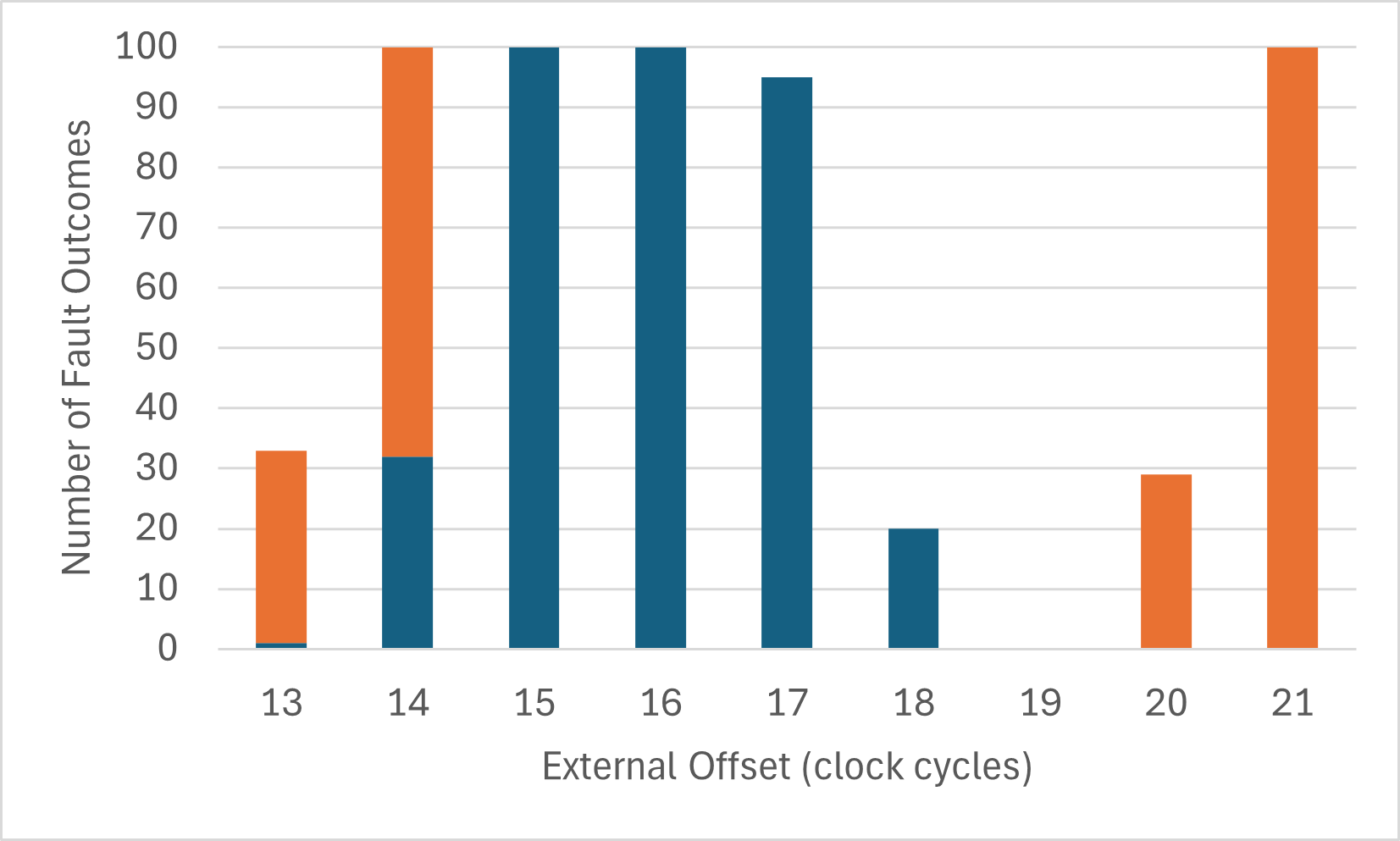}
    \caption{1.75 µs}
    \label{fig:btem_outcomes_1_75}
\end{subfigure}
\hfill
\begin{subfigure}[b]{0.2\textwidth}
    \centering
    \includegraphics[width=\textwidth]{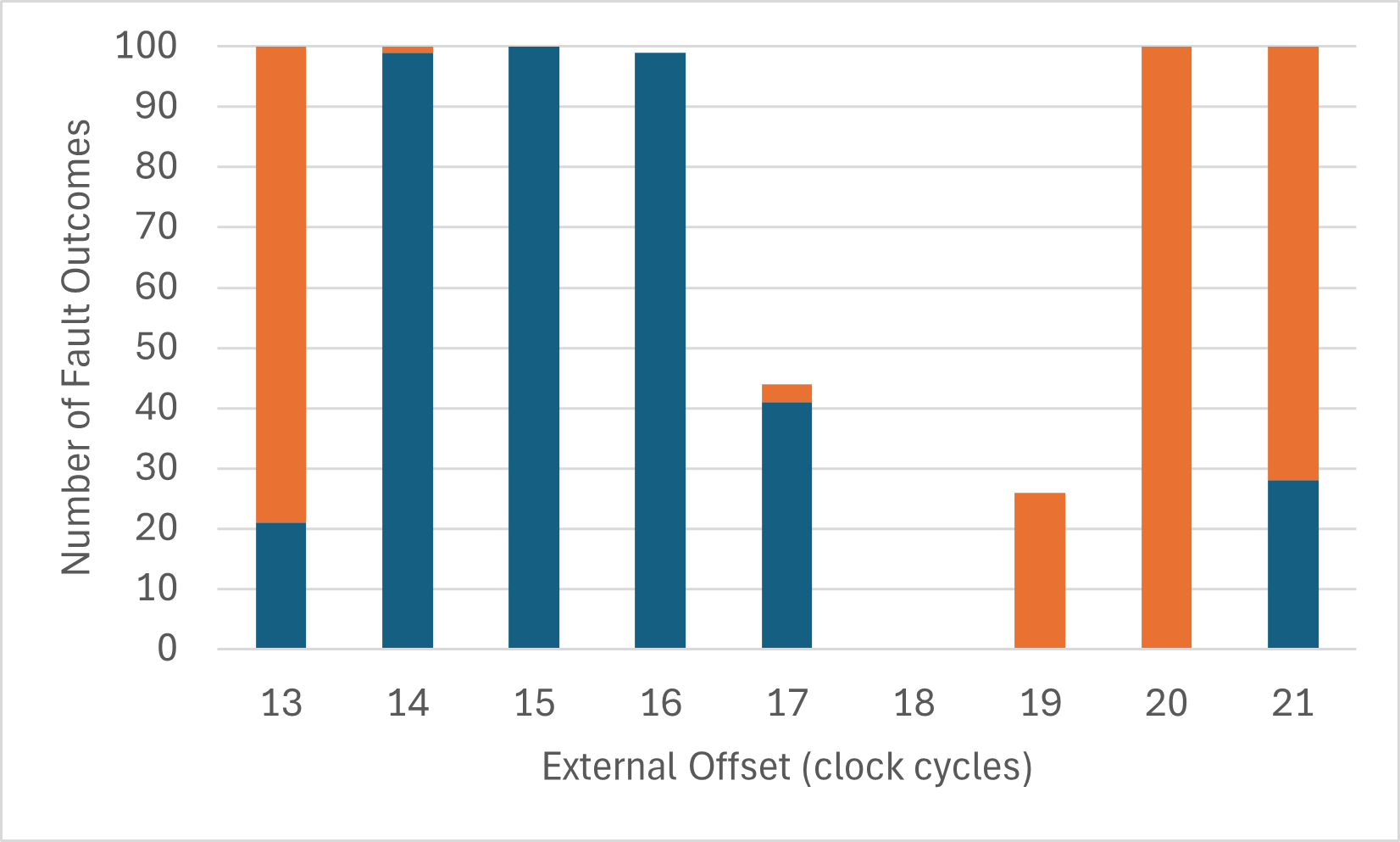}
    \caption{1.78 µs}
    \label{fig:btem_outcomes_1_78}
\end{subfigure}

\caption{Distribution of aggregate fault outcomes (Success, Reset) versus \textit{External Offset}  for Battery Emergency scenario}
\label{fig:btem_outcomes_vs_offset}
\end{figure}

Because this scenario incorporated additional failsafe logic before the target function, temporal alignment required adjustment. Successful physical exploits were not recorded until an \textit{Offset} of 15 clock cycles. As shown in Figure~\ref{fig:btem_outcomes_vs_offset}, when the testing window was narrowed to physical offset of 13 to 21 cycles, a dense cluster of physical faults emerged within a 4-cycle window (in ChipWhisperer timing). Applying the clock ratio, this corresponds to approximately 21 clock cycles within the STM32 framework, logically aligning with the highly vulnerable 45 to 64 clock cycle window identified in the ARMORY simulation. 

\subsubsection{Fault Effect Correlation}
\begin{figure}[ht]
    \centering
    \includegraphics[width=\linewidth]{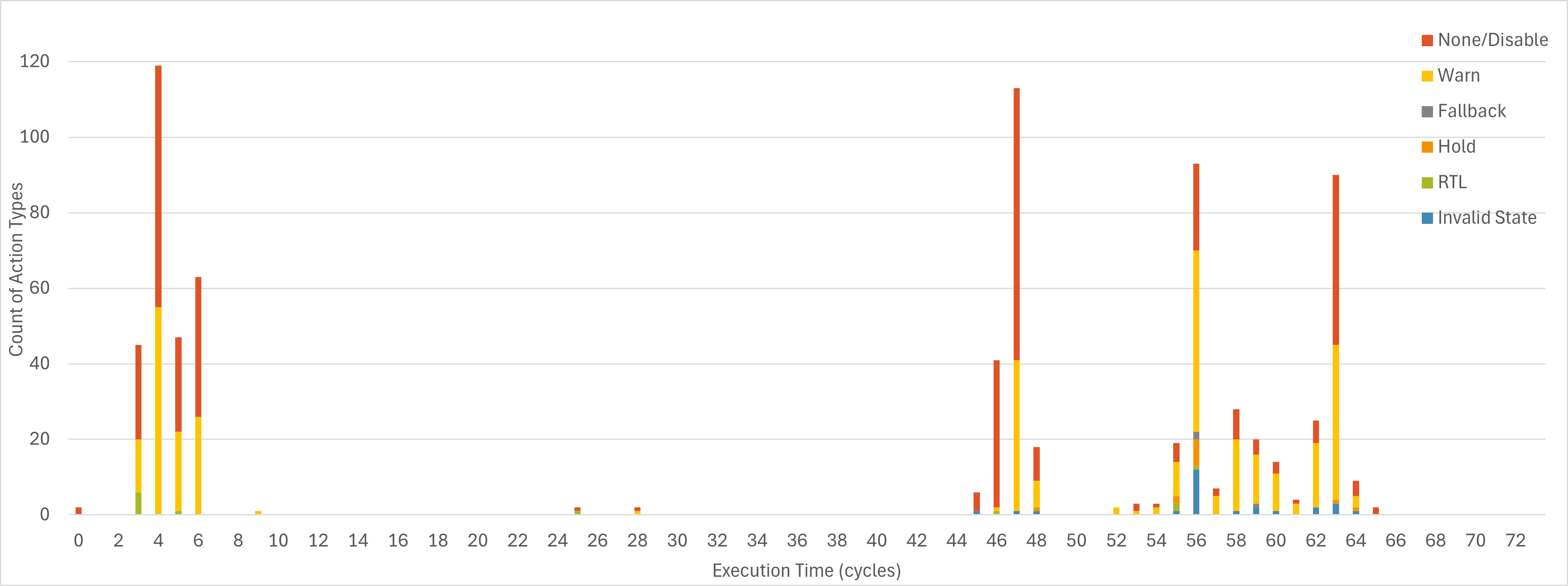}
    \caption{Temporal distribution of fault effects categorized by behavior outcome for Battery Emergency scenario}
    \label{fig:fs_bt_em_action_temporal}
\end{figure}

\begin{figure}[htbp]
\centering
\begin{subfigure}[b]{0.45\textwidth} 
\centering
\includegraphics[width=\textwidth]{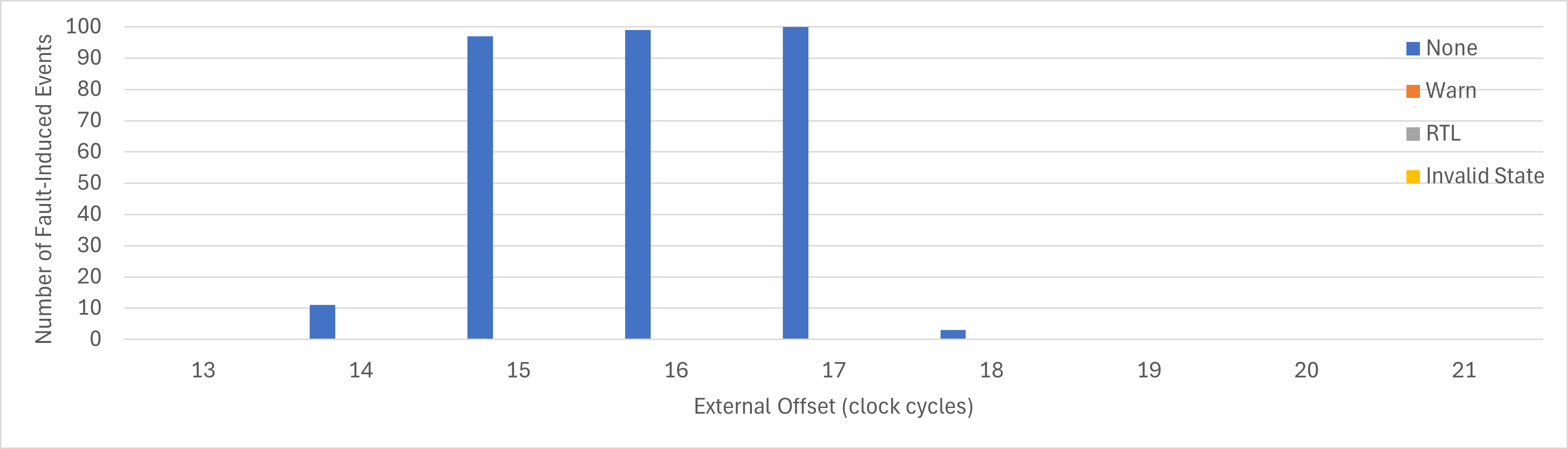} 
\caption{1.69 µs}
\label{fig:btem_action_1_69}
\end{subfigure}
\hfill
\begin{subfigure}[b]{0.45\textwidth} 
\centering
\includegraphics[width=\textwidth]{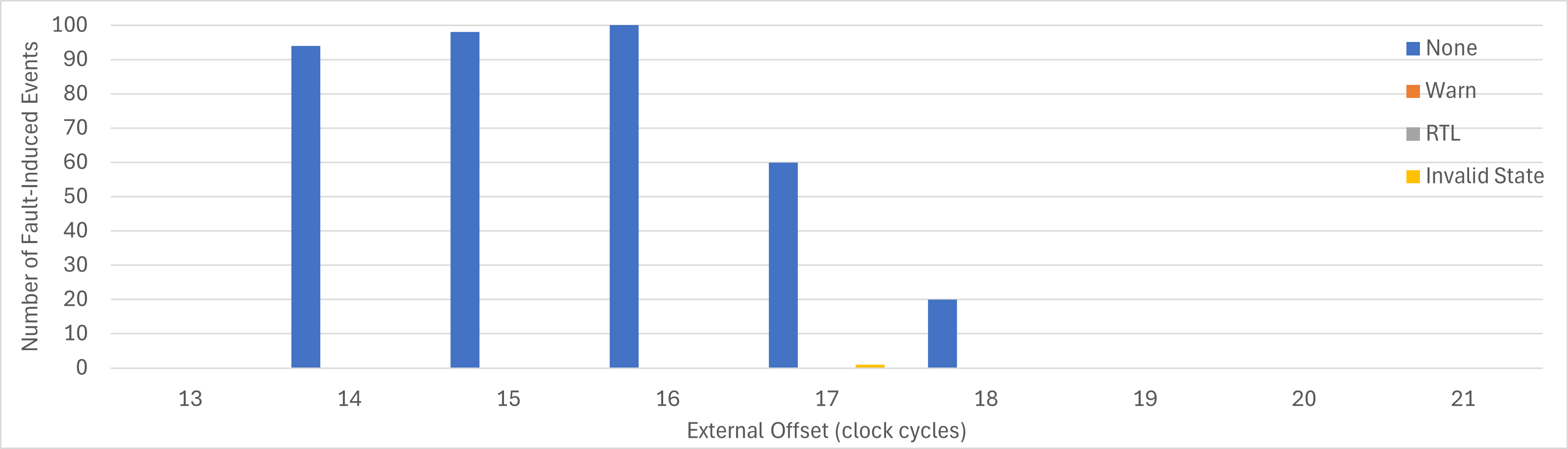}
\caption{1.72 µs}
\label{fig:btem_action_1_72}
\end{subfigure}

\vspace{1em}

\begin{subfigure}[b]{0.45\textwidth} 
    \centering
    \includegraphics[width=\textwidth]{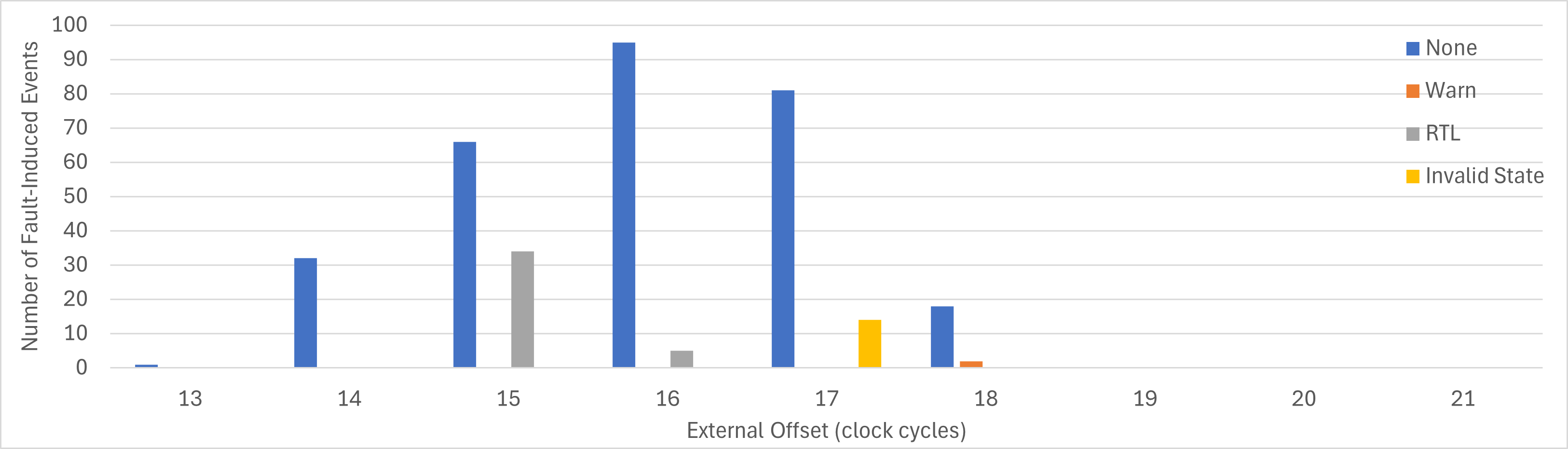}
    \caption{1.75 µs}
    \label{fig:btem_action_1_75}
\end{subfigure}
\hfill
\begin{subfigure}[b]{0.45\textwidth} 
    \centering
    \includegraphics[width=\textwidth]{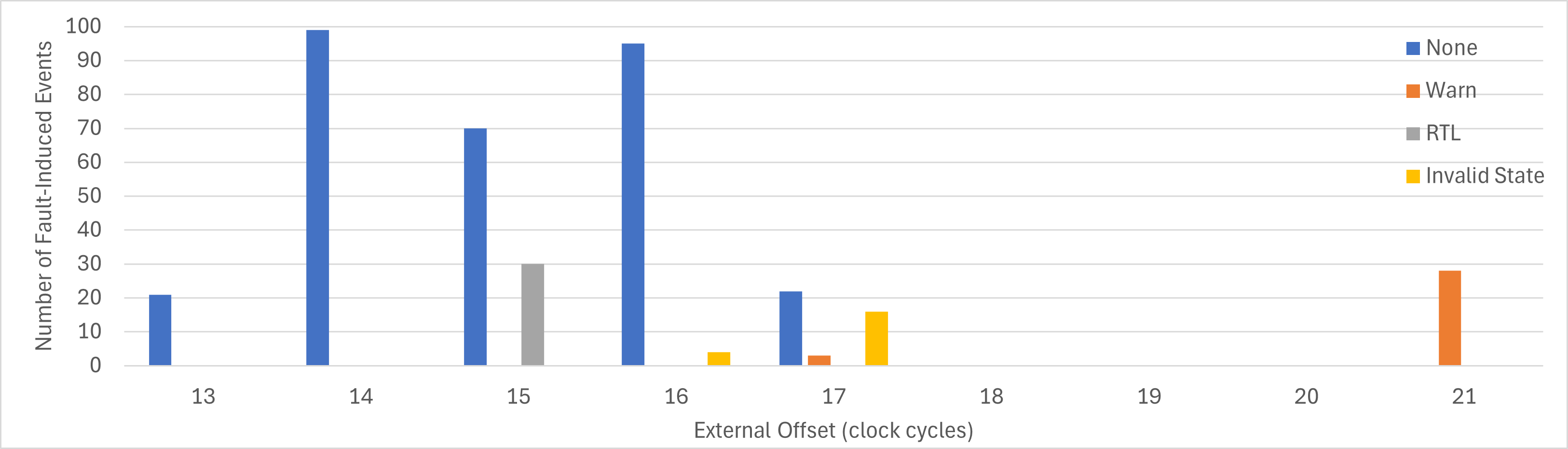} 
    \caption{1.78 µs}
    \label{fig:btem_action_1_78}
\end{subfigure}
\caption{Distribution of specific observed outcomes versus \textit{External Offset}  for Battery Emergency scenario}
\label{fig:btem_action_outcomes_vs_offset}
\end{figure}

Evaluating the severity of the exploits (Figure~\ref{fig:fs_bt_em_action_temporal}), the most frequently observed physical responses were \textit{None/Disable} and \textit{Warn}. This indicates that inducing a fault at almost any vulnerable point within the helper logic successfully degrades the system's ability to trigger the emergency landing.  Furthermore, as detailed in the hardware experiment results (Figure~\ref{fig:btem_action_outcomes_vs_offset}), the most prominent fault-induced outcome was \textit{None}. Forcing the flight controller into a state of total inaction during an emergency battery level proved to be the most reliable and devastating exploit in this scenario.

\subsection{Voltage Glitch Effect on PX4 System Behavior}
\begin{table}[h]
\centering
\caption{Simulation Result of the Voltage Glitch's Effect}
\label{tab:failsafe-actions}
\scalebox{1.1}{
\begin{tabular}{|c|c|c|} 
\hline
\multicolumn{1}{|c|}{\textbf{Scenario}} & \multicolumn{1}{c|}{\textbf{Glitched Behavior}} & \multicolumn{1}{c|}{\textbf{Expected Behavior}} \\
\hline
RC Signal Loss & Hold & Return to Launch\\
Battery Critical & Hold & Return-to-Launch\\
Battery Emergency & Land & Land\\
\hline
\end{tabular}
}
\end{table}

We evaluate the effect of a successful voltage glitch on the PX4 system in software-in-the-loop simulation using the Gazebo Classic simulator. To emulate the effect of a successful voltage glitch, we manually force the selected failsafe Action to None, thereby discarding the behavior chosen by the failsafe logic. Under the RC signal loss and battery critical scenarios, the pseudo glitch successfully disrupts the expected RTL behavior. However, in the battery emergency scenario, the glitch does not prevent the system from landing, suggesting that the landing control logic has higher priority than RTL in the PX4 system.

%% file: sections/conclusion.tex
\section{Conclusion}
This paper presented a novel, highly localized physical threat model targeting Unmanned Aerial Vehicles (UAVs): an autonomous, discrete voltage glitch implant designed to bypass critical flight controller failsafes. To validate the viability of this threat, we employed a rigorous two-phase methodology, correlating machine-code level vulnerabilities identified via ARMORY software simulation with empirical fault injection data gathered using a ChipWhisperer device to attack an STM32-based flight controller board.

Our evaluation across three critical PX4 failsafe scenarios demonstrated the severe practicality of this attack vector. The findings revealed timing-sensitive vulnerabilities where precisely injected faults reliably corrupted critical instruction execution within the failsafe helper logic. Most alarmingly, the physical experiments revealed a strong tendency for the hardware to completely bypass intended safety protocols. Furthermore, these physical experiments exposed the system’s susceptibility to extended transient faults that triggered HardFault exceptions; complex physical behaviors not observable through ARMORY alone, underscoring the absolute necessity of empirical hardware-based analysis.

Ultimately, the ability to neutralize a UAV's final line of defense highlights a critical blind spot in current flight controller architectures, demonstrating that future mitigation strategies must embrace hardware-software co-design to secure the underlying processor.